\documentclass[twocolumn,preprintnumbers,amsmath,amssymb,aps,pra]{revtex4-2}
\usepackage{mathtools}
\usepackage{gensymb}
\usepackage[utf8]{inputenc}
\usepackage{graphicx}
\usepackage[tight,TABTOPCAP]{subfigure}
\usepackage{kbordermatrix}
\renewcommand{\kbldelim}{(}
\renewcommand{\kbrdelim}{)}

\usepackage{tikz}

\usepackage{dcolumn}
\newcolumntype{C}{>{$}c<{$}}
\newcolumntype{R}{>{$}r<{$}}
\newcolumntype{d}[1]{D{.}{.}{#1}}

\usepackage{booktabs}
\usepackage{color}

\newcommand*{\mat}[1]{\boldsymbol{#1}}

\newcommand*{\coord}[1]{\mathbf{#1}}
\newcommand*{\vecr}{\coord{r}}
\newcommand*{\vecx}{\coord{x}}

\newcommand*{\crea}[1]{\hat{#1}^{\dagger}}
\newcommand*{\anni}[1]{\hat{#1}^{\vphantom{\dagger}}}

\DeclareMathOperator*{\argmin}{arg\,min}
\DeclareMathOperator{\trace}{tr}
\DeclareMathOperator{\Trace}{Tr}

\DeclarePairedDelimiter{\abs}{\lvert}{\rvert}
\DeclarePairedDelimiterX\braket[2]{\langle}{\rangle}{#1\delimsize\vert#2}
\DeclarePairedDelimiterX\brakket[3]{\langle}{\rangle}{#1\delimsize\vert#2\delimsize\vert#3}
\DeclarePairedDelimiterX\set[2]{\lbrace}{\rbrace}{#1 : #2}

\newcommand*{\group}{G}

\newcommand*{\groupRDMs}{\mathcal{P}^{N,\group}}

\newcommand*{\vGroupRDMs}{\mathcal{V}^{N,\group}}

\newcommand*{\e}{\mathrm{e}}
\newcommand*{\eqspace}{\phantom{{} = {}}}
\newcommand*{\half}{\frac{1}{2}}
\newcommand*{\im}{\textrm{i}}
\newcommand*{\isDefinedAs}{\coloneqq}
\newcommand*{\thalf}{\tfrac{1}{2}}
\newcommand*{\ud}{\textrm{d}}

\allowdisplaybreaks[4]

\definecolor{DarkBlue}{rgb}{0.2,0.2,0.5}
\definecolor{DarkGreen}{rgb}{0,0.4,0}
\definecolor{Indigo}{rgb}{0.51,0,0.51}

\usepackage[pdftitle={\@title}, pdfauthor={\@author}, colorlinks=true, unicode, bookmarksnumbered=true, linkcolor=DarkBlue, citecolor=DarkGreen, urlcolor=Indigo]{hyperref}

\begin{document}

\title{Implications of the unitary invariance and symmetry restrictions on the development of proper approximate one-body reduced density matrix functionals}
\author{K.J.H. Giesbertz}
\affiliation{Department of Theoretical Chemistry, Vrije Universiteit Amsterdam, De Boelelaan 1083, 1081 HV Amsterdam, The Netherlands}
\email{k.j.h.giesbertz@vu.nl}

\date{\today}

\begin{abstract}
In many of the approximate functionals in one-body reduced density matrix (1RDM) functional theory, the approximate two-body reduced density matrix (2RDM) in the natural orbital representation only depends on the natural occupation numbers.
In \href{http://dx.doi.org/10.1103/PhysRevA.92.012520}{Phys.\ Rev.\ A~\textbf{92}, 012520 (2015)} Wang and Knowles initialised a discussion of to what extent this simplification is valid, by introducing two different H$_4$ geometries with identical natural occupation numbers, but different 2RDMs.
Gritsenko has argued that this feature is due symmetry~[\href{http://dx.doi.org/10.1103/PhysRevA.97.026501}{Phys.\ Rev.\ A~\textbf{97}, 026501 (2018)}].
This work aims to contribute to the discussion on the following points: 
1) one should rather speak of symmetry-restricted variants of the universal functional, than saying that the universal functional is symmetry dependent;
2) the unitary invariance of degenerate NOs can lead to large deviations in the 2RDM elements, especially the phase of the NOs;
3) symmetry-restricted functionals are constructed for the H$_4$ geometries considered by Wang and Knowles, whose structure could serve as guide in the construction of approximate 1RDM functionals.
\end{abstract}

\maketitle

\section{Introduction}

In Ref.~\cite{WangKnowles2015} Wang and Knowles have given an example in which two one-body reduced density matrices (1RDMs) with identical natural occupation spectra correspond to two different ground state two-body reduced density matrices (2RDMs). This is a feature that most common approximate 1RDM functionals cannot handle, since they generate an approximate 2RDM via an explicit algebraic expression only depending on the occupation numbers.
Wang and Knowles argued that a functional dependence on the natural orbitals (eigenfunctions of the 1RDM) also needs to be included in the approximations, if approximate 1RDM functionals aim to handle these isospectral cases correctly.
This feature has been discussed in the context of the symmetry of these systems~\cite{Gritsenko2018, GritsenkoWangKnowles2019}, but symmetry is more a facilitator than the essential factor~\cite{WangKnowles2018}, as will also be stressed in this work.

Though a symmetry-restricted functional is only exact for systems with the prescribed symmetry and generally introduces discontinuities in the potential surface when the symmetry is broken, the idea of imposing symmetry restrictions on the universal functional can still be useful. One advantage is that the lowest excited state energies in each irreducible representation (irrep) can be calculated~\cite{GunnarssonLundqvist1976, Barth1979, Gritsenko2018, GritsenkoWangKnowles2019}. The more important advantage for the current discussion is that the variational freedom in the constrained-search formulation is significantly reduced. This allows one to build explicit parametrisations of these functionals for simple systems, which can serve as a guide for the construction of approximate functionals. In this article I will construct two different symmetry-restricted functionals valid for the square H$_4$ system and H$_2 + 2$H in a minimal basis; the systems used in the demonstration by Wang and Knowles~\cite{WangKnowles2015}. Both functionals are symmetry restrictions of the exact functional, though still flexible enough to deal with both H$_4$ systems.

The article is organised as follows. In Sec.~\ref{sec:exactFuncs}, different formulations of the exact universal functional are discussed and the difference between my definition of `exact' and its use in Refs~\cite{Gritsenko2018, GritsenkoWangKnowles2019} is highlighted. Further, I will show how the results by Wang and Knowles~\cite{WangKnowles2015} can directly be rationalised from the constrained-search formulation. In Sec.~\ref{sec:symAdapt}, symmetry restrictions of the universal 1RDM functional are discussed and it is argued that the true universal functional cannot be considered to be symmetry dependent. In Sec.~\ref{sec:paramD2h}, the two relevant components (irreducible representations) of the symmetry-restricted $D_{2h}$ functional are constructed for the H$_4$ systems in a minimal basis. 
In Sec.~\ref{sec:optimisation}, I investigate how the symmetry-restricted $D_{2h}$ functional operates on the rhombic H$_4$ systems considered in~\cite{GritsenkoWangKnowles2019} 
and discuss the implications of unitary invariance of degenerate natural orbitals (NOs).
In Sec.~\ref{sec:conclusion}, I finalise with conclusions.

\section{Exact universal functionals}
\label{sec:exactFuncs}

There are two different exact interaction energy universal functionals which are useful to consider. Both are exact in the sense that both yield the exact, i.e.\ full configuration interaction, ground state interaction energy within the given (possibly finite) basis we have chosen to work with and both are universal in the sense that they are valid for any non-local one-body potential in that basis.
It is important to realise that a narrower definition of `exact' might be used, in which only the functional for the Coulomb interaction in the complete basis is deemed exact as in Refs~\cite{Gritsenko2018, GritsenkoWangKnowles2019} seems to be intended, although such a view does not appreciate the generality of the framework of 1RDM functional theory. In particular, it neglects the fact that 1RDM functional theory can be well defined in any finite basis where `exact' corresponds to the full CI solution, as opposed to density functional theory (DFT)~\cite{GiesbertzRuggenthaler2019}.

The difference between these functionals lies in their mathematical properties. Most notable are their different domains and convexity properties.
The first exact universal 1RDM functional useful for our purposes was proposed by Levy~\cite{Levy1979}
\begin{equation}\label{eq:LevySearch}
W_{\text{L}}[\mat{\gamma}] \isDefinedAs \min_{\Psi \to \mat{\gamma}}\brakket{\Psi}{\hat{W}}{\Psi} ,
\end{equation}
where $\hat{W}$ denotes the interaction operator. The constrained search only runs over pure states, which has the disadvantage that its domain is difficult to characterise, i.e.\ the so-called pure state $N$-representable 1RDMs~\cite{BorlandDennis1970, Klyachko2006, AltunbulakKlyachko2008}. To this end, Valone proposed to extend the search over mixed states~\cite{Valone1980a}
\begin{equation}\label{eq:ValoneSearch}
W_{\text{V}}[\mat{\gamma}] \isDefinedAs \min_{\hat{\rho} \to \mat{\gamma}}\Trace\bigl\{\hat{\rho}\,\hat{W}\bigr\} ,
\end{equation}
where $\hat{\rho}$ denotes the (full) density-matrix operator and the trace runs over the full Hilbert / Fock space. The advantage is that the domain is now the enlarged and more convenient set of ensemble $N$-representable 1RDMs~\cite{Coleman1963}. An additional advantage is that this functional is convex by construction~\cite{PhD-Giesbertz2010, PhD-Baldsiefen2012, Schilling2018, GiesbertzRuggenthaler2019}, which guarantees that any minimum found during minimisation over $\mat{\gamma}$ will be global.

Irrespective of which functional we use, we can express these universal functionals via the 2RDM as
\begin{equation}\label{eq:2RDMSearch}
W[\mat{\gamma}] = \half\min_{\mat{\Gamma} \to \mat{\gamma}}\sum_{ijkl}\Gamma_{ij,kl}\braket{ij}{kl} ,
\end{equation}
where we use the same notation as Wang and Knowles~\cite{WangKnowles2015}, i.e., $\braket{ij}{kl}$ are the two-electron integrals in physicist notation and the 2RDM is defined as $\Gamma_{ij,kl} = \brakket{\Psi}{\crea{a}_i\crea{a}_j\anni{a}_l\anni{a}_k}{\Psi}$ for pure states and $\Gamma_{ij,kl} = \trace\bigl\{\hat{\rho}\,\crea{a}_i\crea{a}_j\anni{a}_l\anni{a}_k\bigr\}$ for the more general mixed states.
Note that the one-particle basis to which the indices of the 2RDM refer, is implied by the two-electron integrals, on which the functional implicitly depends. The dependence of the functional on the two-electron integrals has been studied in more detail recently~\cite{CioslowskiMihalkaSzabados2019, Cioslowski2020}.
The minimisation should only search over pure (Levy) or ensemble (Valone) $N$-representable 2RDMs, to ensure that a corresponding pure or mixed state exists which yields this 2RDM and hence, that the variational principle applies.

Any one-particle basis can be used in this form of the universal functional. We can use this freedom to simplify the constraint on the 2RDM by working in the natural orbital (NO) basis of the requested 1RDM~\cite{
CioslowskiPernalZiesche2002, KollmarHess2003, CioslowskiBuchowieckiZiesche2003, KollmarHess2004}, so the NOs now implicitly enter via the two-electron integrals. As the 1RDM is diagonal by definition, the constraint on the 2RDM reduces to
\begin{equation}\label{eq:2RDMtoOccMap}
\sum_j\Gamma_{ij,kj} = (N-1)n_i\delta_{ik} ,
\end{equation}
where $n_i$ are the natural occupation numbers, i.e.\ eigenvalues of the 1RDM.

This means that if we transform the two-electron integrals to the NO-basis, the constraint $\mat{\Gamma} \to \mat{\gamma}$ now only needs to consider the occupation numbers
\begin{equation}\label{eq:NOconstrSearch}
W[\mat{\gamma}] = \half\min_{\mat{\Gamma} \to \mat{n}}\sum_{ijkl}\Gamma_{ij,kl}\braket{ij}{kl}_{\text{NO}} .
\end{equation}
Since the mapping $\mat{\Gamma} \mapsto \mat{n}$ in~\eqref{eq:2RDMtoOccMap} is many-to-one, there will be remaining degrees of freedom in~\eqref{eq:NOconstrSearch} over which the minimisation should be performed. It is convenient to make these remaining degrees of freedom explicit by using a parametrisation $\mat{\xi}$ for the set of $N$-representable 2RDMs yielding the requested occupation number spectrum $\overline{\Gamma}_{ij,kl}[\mat{\xi},\mat{n}]$. The constrained search in~\eqref{eq:NOconstrSearch} can now be expressed as an unconstrained minimisation over the remaining degrees of freedom $\mat{\xi}$~\cite{PernalCioslowski2005}
\begin{equation}\label{eq:freeSearch}
W[\mat{\gamma}] = \half\min_{\mat{\xi}}\sum_{ijkl}\overline{\Gamma}_{ij,kl}[\mat{\xi},\mat{n}]\,\braket{ij}{kl}_{\text{NO}} .
\end{equation}
Note that the parametrisation $\overline{\Gamma}_{ij,kl}[\mat{\xi},\mat{n}]$ only needs to be concerned about the occupation numbers and \emph{not the NOs}, since $N$-representability does not depend on the orbital basis~\cite{Coleman1963}. 
It should be obvious that such a parametrisation in terms of $\mat{\xi}$ is definitely not unique. There are only convenient and less convenient parametrisations, depending on the situation.

In principle, the route towards construction of such an explicit functional is straightforward: Given the 1RDM $\gamma$, 1) write down a parametrisation of the wavefunction or the density-matrix operator in the NO basis, 2) eliminate parameters that are determined by the 1RDM constraint 3) contract the wavefunction / density-matrix operator to the 2RDM and 4) obtain the functional value by optimisation of the remaining parameters. Unfortunately, the 1RDM constraint enters in a non-linear way, which makes the elimination of parameters a non-trivial task in practice. Especially the positivity constraint on the density-matrix operator is difficult to conciliate with the 1RDM constraint. A parametrisation for the Valone functional~\eqref{eq:ValoneSearch} has therefore only been explicitly constructed for the two-site Hubbard model~\cite{Schilling2018} and Anderson model~\cite{TowsPastor2011}.

Working with a constrained search over only pure states~\eqref{eq:LevySearch} is more convenient for an explicit construction. Such an explicit construction is readily possible for the two-electron case, thanks to the Schmidt decomposition~\cite{Schmidt1907} (sometimes referred to as the Carlson--Keller expansion~\cite{CarlsonKeller1961}), which makes the constraint $\Psi \to \mat {\gamma}$ trivial. The anti-symmetry additionally requires the NOs $\phi_k(\vecx)$ to be pairwise degenerate~\cite{Coleman1963}, which we express by using positive and negative indices as: $n_k = n_{-k}$. The wavefunction can now be expressed as~\cite{CioslowskiPernalZiesche2002, PhD-Giesbertz2010, RappBricsBauer2014}
\begin{equation}\label{eq:twoElecWave}
\Psi(\vecx_1,\vecx_2)
= \sum_{\mathclap{k=1}}^M\sqrt{n_k}\e^{\im\xi_k}\,\abs[\big]{\phi_k(\vecx_1)\phi_{-k}(\vecx_2)} ,
\end{equation}
where $\vecx = \vecr\sigma$ is a combined space-spin coordinate and $2M$ is the number of spin-orbitals in the basis~\footnote{In the case of an odd number of spin-orbitals, one of the spin-orbitals drops out of the expansion, so has $n_0 = 0$.}. The 2RDM is readily found to be
\begin{equation}\label{eq:exactTwoRDM}
\overline{\Gamma}^{\text{2el}}_{ij,kl}[\mat{\xi},\mat{n}]
= \sqrt{n_in_k}\e^{\im(\xi_k - \xi_i)}\delta_{i,-j}\delta_{k,-l} ,
\end{equation}
with $\xi_{-k} = \xi_k + \pi$. The free parameters $\mat{\xi}$ in the exact two-electron functional are the phases in the two-electron wavefunction~\eqref{eq:twoElecWave}, which are the only degrees of freedom not fixed by the 1RDM.

Wang and Knowles actually seem to dismiss such a form of the exact functional, which internally houses an additional variable set $\mat{\xi}$: ``One may take the phase as additional variables~\cite{GiesbertzGritsenkoBaerends2010a}, then this will go beyond Gilbert’s original variable set: the natural orbitals and their occupation numbers.''
One has to keep in mind, however, that all density-functional-like theories are actually reformulations of the Schrödinger equation. This means that the full flexibility of the complete many-body state cannot magically disappear and needs to be accounted for somewhere in the theory. Levy's constrained-search formulation makes this very explicit by minimising over all pure states~\eqref{eq:LevySearch}. The extension by Valone even extends the search to mixed states~\eqref{eq:ValoneSearch}. This additional variational freedom beyond degrees of freedom of the 1RDM persists of course when we reformulate the exact functional as a search over $N$-representable 2RDMs~\eqref{eq:2RDMSearch} and is made explicit as the variable set $\mat{\xi}$ in~\eqref{eq:freeSearch}.

The many-to-one relation in the mapping $\mat{\Gamma} \to \mat{n}$
is actually the crucial property to explain the results presented by Wang and Knowles in Ref.~\cite{WangKnowles2015}. Though the occupation numbers can be made identical in both systems (square H$_4$ and H$_2 + 2$H), the NOs are different and hence, lead to a different set of two-electron integrals in~\eqref{eq:freeSearch}. As the two-electron integrals have different values, the minimisation over the variables $\mat{\xi}$ will lead to a different minimum and thus, a different 2RDM.

It might be that Wang and Knowles have a different functional form of the 2RDM in mind. Since every observable can be regarded as a functional of the 1RDM in 1RDM functional theory~\footnote{We only know this for pure $v$-representable states~\cite{Gilbert1975} to be the case and also in general at elevated temperature~\cite{GiesbertzRuggenthaler2019}},
also the ground state 2RDM is a functional of the 1RDM, or equivalently, a functional of the NOs and occupation numbers $\mat{\Gamma}^{\text{NO}}[\{\phi\},\mat{n}]$. This is the functional aimed for by many approximate functional developers, since in that case the interaction energy is directly given as
\begin{equation}
W[\{\phi\},\mat{n}] = \half\sum_{ijkl}\Gamma^{\text{NO}}[\{\phi\},\mat{n}]_{ij,kl}\braket{ij}{kl}_{\text{NO}} ,
\end{equation}
which does not contain an internal optimisation. The disadvantage is that the functional dependence of $\mat{\Gamma}^{\text{NO}}[\{\phi\},\mat{n}]$ is more complicated than $\overline{\mat{\Gamma}}[\mat{\xi},\mat{n}]$: even for simple systems, no explicit form of $\mat{\Gamma}^{\text{NO}}[\{\phi\},\mat{n}]$ is known%
~\footnote{Another complication for degenerate states is that $\Gamma^{\text{NO}}[\{\phi\},\mat{n}]_{ij,kl}$ is not unique anymore}.
Both functionals are related as
\begin{align}\label{eq:GammaRel}
\mat{\Gamma}^{\text{NO}}[\{\phi\},\mat{n}]
&= \argmin_{\mat{\Gamma} \to \mat{n}}\sum_{ijkl}\Gamma_{ij,kl}\braket{ij}{kl}_{\text{NO}} \notag \\*
&= \overline{\mat{\Gamma}}[\mat{\xi}_{\text{opt}},\mat{n}] ,
\end{align}
where $\mat{\xi}_{\text{opt}}$ are the optimal parameters in~\eqref{eq:freeSearch}.
So the NO dependent variant can be obtained by performing the minimisation over the parameters $\mat{\xi}$ in~\eqref{eq:freeSearch} [or the constrained search in~\eqref{eq:NOconstrSearch}] and then the 2RDM elements can be extracted.
This implicit dependence of $\mat{\Gamma}^{\text{NO}}[\{\phi\},\mat{n}]$ on the NOs makes it inconvenient to build approximations which try to capture this NO dependence directly.
The construction of approximate $\overline{\Gamma}_{ij,kl}[\mat{\xi},\mat{n}]$ is more feasible and therefore provides a better starting point for approximate 1RDM functionals that aim to go beyond a simple dependence on the natural occupation numbers.

It is worth pointing out that relation~\eqref{eq:GammaRel} implies that the functional dependence on the NOs of the 2RDM only vanishes, if $\overline{\Gamma}_{ij,kl}$ also does not depend on auxiliary parameters. In this case, these functionals are even equal
\begin{equation}
\Gamma^{\text{NO}}_{ij,kl}[\mat{n}]
= \overline{\Gamma}_{ij,kl}[\mat{n}] .
\end{equation}
However, such a simple form can never occur for the exact functional, except in very limited settings, e.g.\ two electrons in two orbitals.
In Ref.~\cite{GritsenkoWangKnowles2019} it was asserted that also the exact constructions for two-electron systems~\cite{LowdinShull1956} and for translationally invariant one-band lattice models~\cite{SchillingSchilling2019} are of this simple form $\Gamma_{ij,kl}[\mat{n}]$, but that is incorrect.
For the two-electron system, we have an internal optimisation over phase factors~\eqref{eq:exactTwoRDM} and translationally invariant one-band lattice models the internal minimisation is actually clearly mentioned just after Eq.~(9) in Ref.~\cite{SchillingSchilling2019}. Only with additional assumptions, can the parameters $\mat{\xi}$ be eliminated. An exact functional for the rhombus H$_4$ system in Ref.~\cite{GritsenkoWangKnowles2019} is given in Sec.~\ref{sec:paramD2h}, but will clearly not be of the simple form $\overline{\Gamma}_{ij,kl}[\mat{n}]$.
We will see later in Sec.~\ref{sec:optimisation} that variations in the phase of the NOs can lead to significant deviations in the 2RDM elements.

\section{Symmetry restrictions}
\label{sec:symAdapt}

The pure-state expression for the 2RDM of two-electron systems~\eqref{eq:exactTwoRDM} is completely general, valid for any spin-dependent potential and spin-dependent two-electron interaction. However, often we work with the spin-independent Coulomb interaction and spin-independent potentials (no magnetic fields). This means that the Hamiltonian commutes with the spin-operators and the eigenstates can be classified according to their spin-state. The constrained search can therefore be restricted to either singlet or triplet states as originally done by Löwdin and Shull~\cite{LowdinShull1956}. They additionally used that the Hamiltonian is now also real, so the eigenstates can be chosen to be real and hence, the phase factors $\e^{\im\xi_k} = \pm 1$.

Let us put this in a more general setting.
In case we are only interested in external potentials / Hamiltonians with a certain symmetry, we know that the ground state will belong to one of the irreducible representations (irreps) of the symmetry group. The constrained search can therefore be broken down into separate constrained searches over each irrep $I$ of the symmetry group $\group$~\footnote{We could also extend the definition~\eqref{eq:symAdapIrrep} to ensembles over states of the specific irrep $I$, but this additional flexibility does not seem to be useful for this discussion.}
\begin{subequations}\label{eq:symAdapt}
\begin{align}\label{eq:symRestrFunc}
W^{\group}[\mat{\gamma}]
&\isDefinedAs \min_{I} W^G_I[\mat{\gamma}] ,
\intertext{where}
\label{eq:symAdapIrrep}
W^{\group}_I[\mat{\gamma}] &\isDefinedAs \inf_{\mathclap{\Psi^{\group,I} \to \mat{\gamma}}}\brakket{\Psi^{\group,I}}{\hat{W}} {\Psi^{\group,I}} \\
&= \half\inf_{\mat{\xi}}\sum_{ijkl}\overline{\Gamma}^G_{ij,kl}[I, \mat{\xi},\mat{n}]\,\braket{ij}{kl}_{\text{NO}}. \notag
\end{align}
\end{subequations}
In the definition of $W^{\group}_I[\mat{\gamma}]$ we have put an infimum, since it is well possible that no $\Psi^{\group,I} \to \mat{\gamma}$ can be found, in which case we set $W^{\group}_I[\mat{\gamma} \nleftarrow \Psi^I] = +\infty$.
Since we retained the minimum in~\eqref{eq:symRestrFunc}, we assume that we only allow for 1RDMs which can be generated by a wavefunction of one of the irreps of the group $\group$. One could call these pure-state $N,\group$-representable 1RDMs
\begin{equation}
\groupRDMs = \set{\gamma}{\exists \Psi^{N,\group} \to \gamma} .
\end{equation}
By restricting ourselves to all potentials which have the symmetry $\group$, $W^{\group}[\mat{\gamma}]$ is an exact functional in the sense that for all 1RDMs which come from a ground-state wavefunction of that symmetry group, it will yield the exact ground state interaction energy: the $v^{\group}$-representable 1RDMs
\begin{equation}
\vGroupRDMs = \set{\gamma}{\exists v^{\group} \to \Psi_{\text{g.s.}}^{N,\group} \to \gamma} .
\end{equation}
So this can be regarded as a restriction in the universality of the theory, with a corresponding adaptation of the domains.

This is basically the idea that has been put forward by Gritsenko~\cite{Gritsenko2018, GritsenkoWangKnowles2019} to explain that different ground state 2RDMs could correspond to identical occupation number spectra, in which he refers to symmetry dependence of the exact functional.
However, from the preceding discussion it follows that for clarity one should rather not talk about a single functional, but about a \emph{set} of functionals, each of them valid for a particular symmetry group $\group$.
Since each of these functionals is only exact for potentials exhibiting that particular symmetry (and corresponding $v^{\group}$-representable 1RDMs), I rather like to stress that we have \emph{restricted} the universality of the ``parent'' functional~\eqref{eq:LevySearch}, than saying that the exact functional is symmetry dependent. In other words, for each symmetry group we can construct a separate 1RDM functional theory for that group, each with its own (simplified) symmetry-restricted functional~\eqref{eq:symRestrFunc}.

An obvious relation between these symmetry-restricted 1RDM functional theories is that the 1RDM functional theory of a group $\group$ is effectively contained in the 1RDM functional theory of a subgroup $\group'$ of $\group$, because potentials of symmetry $\group$ also belong to the subgroup $\group'$ by definition. So ultimately the ``no-symmetry'' group contains all symmetry-restricted versions and coincides with the original universal 1RDM functional we started with (Sec.~\ref{sec:exactFuncs}). Hence, a functional for any subgroup $\group_k$ of a group $\group$ can be used as an exact functional for 1RDM functional theory of the group $\group$. Obviously, the converse does not hold, since $W^{\group}[\mat{\gamma}] \geq W^{\group_k}[\mat{\gamma}]$ for $v^{\group_k}$-representable 1RDMs and for all $\mat{\gamma} \notin \vGroupRDMs$ a strict inequality is expected: $W^{\group}[\mat{\gamma}] > W^{\group_k}[\mat{\gamma}]$.
It is therefore not possible to derive general properties of the original universal functional~\eqref{eq:LevySearch} from these symmetry-restricted functionals. Also an attempt to combine the symmetry-restricted functionals by minimising over all of them~\cite{Gritsenko2018} is of no avail, since any system belongs to the ``no-symmetry'' group, which is simply the Levy-constrained-search functional~\eqref{eq:LevySearch}
\begin{equation}\label{eq:symMinIsNoSym}
\min_{\group}W^{\group}[\mat{\gamma}] = W^{\text{no sym}}[\mat{\gamma}] = W_{\text{L}}[\mat{\gamma}] .
\end{equation}
From these considerations it should be clear that it is hard to make any exact statements on the usual exact Levy functional~\eqref{eq:LevySearch} based on symmetry.

But there is no need to invoke any symmetry argument to explain that different ground state 2RDMs can correspond to identical occupation number spectra. We can resort to the simple argument presented in Sec.~\ref{sec:exactFuncs}, which is completely sufficient and does not make any reference to symmetry. The only difference in the symmetry-restricted setting is that it becomes natural to choose the irrep as one of the parameters in the constrained-search, which is what has effectively been done in~\eqref{eq:symRestrFunc}.

So far the discussion is only about exact statements, but practical 1RDM functional theory aims to deliver an approximation which is good enough, but does not need to be exact. In general it should not be exact, otherwise it would be computationally too costly to be of any practical use. This is where the symmetry-restricted variant could play an important role in the development of practical 1RDM functional theory. Even if a system does not exactly belong to a symmetry group $\group$, the functional $W^{\group}$ is still expected to provide a very accurate approximation to the exact value and the corresponding 2RDM matrix elements (in the NO basis). In Ref.~\cite{GritsenkoWangKnowles2019} such an approximation to the exact functional is referred to as a practical functional.

To assess some of these ideas, we will examine the systems studied in~\cite{GritsenkoWangKnowles2019} in more detail. Of particular interest is the sequence of rhombi with a varying apex from $90\degree$ to $120\degree$ and adjusted sides such that the natural occupation numbers remain identical. It has been shown that the eigenvalues of the 2RDM do not vary significantly, which hints that the remaining variational freedom within a single irrep is not very significant. However, in the development of 1RDM functionals we do not work with the eigenvalues of the 2RDM, but its matrix elements in NO basis. So it is better to investigate the dependence of the 2RDM matrix elements directly. We will find for this sequence that indeed the magnitude of these elements does not vary much, but the sign of these elements poses a difficulty.

In this more detailed investigation, we will also use the symmetry-restricted functionals to simplify the parametrisation $\mat{\xi}$ which needs to be established.
Yet, let us first consider the explicit construction of a pure-state functional (with or without symmetry restriction) in more detail.
Assuming that the wavefunction is expanded in NOs, the constrained search~\eqref{eq:LevySearch} imposes two types of conditions: 1) diagonality conditions, i.e.\ that the 1RDM is diagonal and 2) occupation number conditions.
The conditions from the occupation numbers do not contain any cross terms between the configurations, so these conditions lead to a set of linear constraints on the square modulus of the CI coefficients
\begin{equation}
\sum_IA_{kI}\;\abs{c_I}^2 = n_k ,
\end{equation}
where $I$ runs over the configurations and one-particle coupling coefficients~\cite{KnowlesHandy1989} $A_{kI}$ tells how much each configuration $I$ contributes to the natural occupation $n_k$.
This linear set of equations can easily be solved to determine the constraints on the CI coefficients, though the large dimension of the CI space can be problematic in practice.
The null-space of $\mat{A}$ yields the remaining variational freedom. Provided that there are no additional constraints, the null-space yields exactly the parameters $\mat{\xi}$ in~\eqref{eq:freeSearch}, apart from the phase of the CI coefficients. This is exactly how the exact functional for the translationally invariant one-band lattice model was constructed~\cite{SchillingSchilling2019}.

On the other hand, the diagonality conditions do mix different configurations, so these conditions contain products of different CI coefficients. Hence, the presence of these conditions leads to coupled quadratic equations, which are difficult to solve in general, although, for simple cases like the examples put forward by Wang and Knowles, they can still be solved (see Sec.~\ref{sec:paramD2h} and Ap.~\ref{ap:paramC2v}), since there are only a few diagonality constraints to deal with when symmetry restrictions are imposed.

\section{The $D_{2h}$ symmetry-restricted functional for the H$_4$ systems in minimal basis}
\label{sec:paramD2h}

The different perspective on the `exact' functional(s), leads to a somewhat different role of the H$_4$ system in a minimal basis. In the narrower definition, the H$_4$ model serves only as an approximation to the complete basis limit, so the constrained-search functionals~\eqref{eq:LevySearch} and~\eqref{eq:ValoneSearch} would be considered approximate~\cite{GritsenkoWangKnowles2019}. In this work, the H$_4$ system is considered as a valid setting in its own right, because one can perfectly define the exact 1RDM functional as the full CI result.

The most general Levy-type functional for these systems would need to deal with 4 electrons in 8 spin-orbitals, i.e.\ a configuration interaction (CI) expansion of $\binom{8}{4} = 70$ terms. However, when using full spin-symmetry, we only need to deal with 20, 15, 1 configuration state functions for the singlet, triplet and quintet irreps respectively.

Gritsenko considered the H$_2 + 2$H system to be arranged in a trapezoid~\cite{Gritsenko2018}, with the shorter of the two parallel sides being the H$_2$ bond and the longer side tending to infinity. The square H$_4$ system can be regarded as a trapezoid with parallel legs. This allows us to use the $C_{2v}$ symmetry group as a common symmetry group for both systems. In this case both systems have their ground state in the $^1A_1$ irrep, which has 12 terms in its expansion. One can actually construct a parametrisation for this wave function, because there are only two conditions to make the corresponding 1RDM diagonal. The construction of this parametrisation is quite involved and not used in the analysis, so has been deferred to Appendix~\ref{ap:paramC2v}. In this section, we will use a higher symmetry group, which makes the construction less complicated and more instructive to get the general idea.

\begin{figure}[t]
\begin{center}
\begin{tikzpicture}
  \draw (1.5,0) -- node[above=3, midway, inner sep=1] {$R$} (0,1) -- node[above=3, midway, inner sep=1] {$R$} (-1.5,0) 
    -- node[below=3, midway, inner sep=1] {$R$} (0,-1) -- node[below=3, midway, inner sep=1] {$R$} cycle;
  \filldraw (1.5,0) circle (0.05) node[anchor=west] {H$_1$} ;
  \filldraw (-1.5,0) circle (0.05) node[anchor=east] {H$_2$} ;
  \filldraw (0,1) circle (0.05) node[anchor=south] {H$_3$} ;
  \filldraw (0,-1) circle (0.05) node[anchor=north] {H$_4$} ;
  \draw (-1.1,0) node[anchor=west] {$\phi$} arc [start angle=0, end angle=33.69, radius=0.4];
  \draw (-1.1,0) arc [start angle=0, end angle=-33.69, radius=0.4];
  \draw [<->] (3,1) node[anchor=south] {$y$} -- (3,0.5) -- (3.5,0.5) node[anchor= west] {$x$};
\end{tikzpicture}
\caption{The orientation of the H$_4$ system organised as a rhombus w.r.t.\ to the cartesian system. The apex is defined as the angle $\phi$ and all sides have equal length $R$.}
\label{fig:H4orientation}
\end{center}
\end{figure}
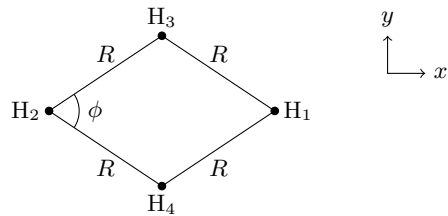

As observed in Ref.~\cite{GritsenkoWangKnowles2019}, the highest common symmetry group is actually $D_{2h}$, if we we arrange the H$_2 + 2$H system in a rhombus (see Fig.~\ref{fig:H4orientation}) instead of a trapezoid. The H$_2$ bond is then placed along the short diagonal and the long diagonal of the rhombus tends to infinity. The square H$_4$ is a rhombus in which the diagonals have equal length. In this case the ground state of the H$_2 + 2$H system belongs to the $^1A_g$ irrep, whereas the ground state of the square H$_4$ system belongs to the $^1B_{1g}$ irrep. We therefore need to construct the symmetry-restricted functional in two irreps, although these are two easier tasks, since the wavefunctions in these irreps only contain 8 and 4 terms respectively. But more importantly, we only need to handle one diagonality constraint.

Let us consider first the simpler $^1B_{1g}$ irrep in detail. A general wavefunction in this irrep can be written as
~\footnote{The last configuration is actually missing in~\cite{GritsenkoWangKnowles2019}, because the authors use the fact that the square H$_4$ has an even higher symmetry $D_{4h}$. In that case the last configuration belongs to the $^1A_{2g}$ irrep and hence, does not contribute to the ground state of square H$_4$. That is probably why the authors in~\cite{GritsenkoWangKnowles2019} label the orbitals by the $D_{4h}$ irreps.}
\begin{align}\label{eq:1B1gWave}
\Psi^{^1B_{1g}} ={}& c_1\abs[\big]{1a_g^2b_{2u}b_{3u}[\alpha\beta - \beta\alpha]} + {} \notag \\
&c_2\abs[\big]{2a_g^2b_{2u}b_{3u}[\alpha\beta - \beta\alpha]} + {} \notag \\
&c_3\abs[\big]{1a_g2a_gb_{2u}b_{3u}[\alpha\beta\alpha\beta + \alpha\beta\beta\alpha + {} \notag \\*
&\qquad
\beta\alpha\alpha\beta + \beta\alpha\beta\alpha - 2(\alpha\alpha\beta\beta + \beta\beta\alpha\alpha)]} + {} \notag \\
&c_4\abs[\big]{1a_g2a_gb_{2u}b_{3u}[\alpha\beta\alpha\beta - \alpha\beta\beta\alpha - {} \notag \\*
&\qquad
\beta\alpha\alpha\beta + \beta\alpha\beta\alpha]} .
\end{align}
This wavefunction can only yield spin-integrated 1RDMs with $n_{b_{2u}} = n_{b_{3u}} = 1$ and $n_{1a_g} + n_{2a_g} = 2$, so only for those 1RDMs $W^{D_{2h}}_{^1B_{1g}} < \infty$.

\begin{figure}[t]
  \includegraphics[width=\columnwidth]{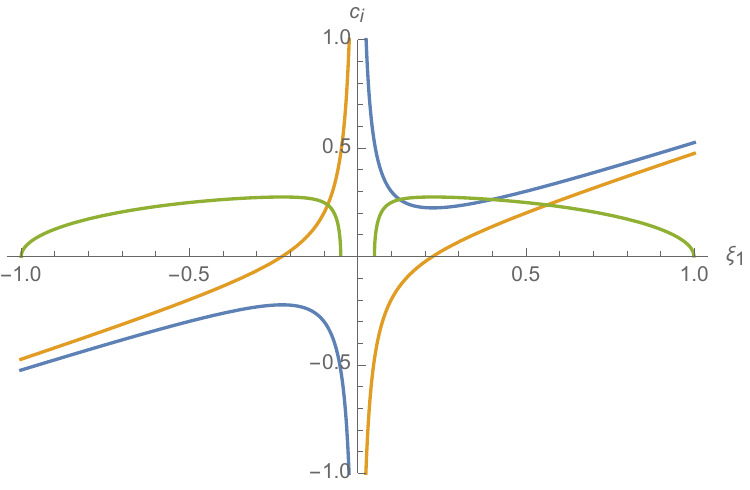}
  \caption{The expansion coefficients of the $^1B_{1g}$ wavefunction~\eqref{eq:1B1gWave} as a function of the variable $\xi_1$ for $n_{1a_g} - n_{2a_g} = 0.2$ as given by~\eqref{eq:B1gCoefs}: $c_1$ (blue), $c_2$ (orange) and $c_3$ (green).}
  \label{fig:B1gcoefs}
\end{figure}

To construct a parametrisation, we first observe that the 1RDM has 2 orbitals in the $a_g$ block, 1 orbital in the $b_{2u}$ block and 1 orbital in the $b_{3u}$ block, so we only need to make the 1RDM diagonal in the $a_g$ block.
The off-diagonal element of the 1RDM from $\Psi^{^1B_{1g}}$ vanishes if $c_4(c_1+c_2) = 0$. Since $c_1+c_2 = 0$ is only possible if $n_{1a_g} = n_{2a_g}$, which is unlikely due to the higher kinetic energy of $n_{2a_g}$, we only parametrise for $c_4 = 0$, which is in agreement with the higher $D_{4h}$ symmetry of the square H$_4$~\cite{WangKnowles2015, GritsenkoWangKnowles2019}. The remaining degrees of freedom can be parametrised with one parameter $\xi_1 = c_1 + c_2$. From the intermediate quantity
\begin{equation}\label{eq:zeta2B1g}
\zeta_2 = c_1 - c_2 = \frac{n_{1a_g} - n_{2a_g}}{4\xi_1} ,
\end{equation}
the CSF coefficients can now be readily calculated as
\begin{subequations}\label{eq:B1gCoefs}
\begin{align}
c_1 &=( \xi_1 + \zeta_2)/2 , \\
c_2 &=( \xi_1 - \zeta_2)/2 , \\
c_3 &= \tfrac{1}{2\sqrt{3}}\bigr[1 - \xi_1^2 - \zeta_2^2 \bigr]^{1/2} .
\end{align}
\end{subequations}
As an illustration, the behaviour of the CI coefficients as a function of $\xi_1$ is shown in Fig.~\ref{fig:B1gcoefs} for $n_{1a_g} - n_{2a_g} = 0.2$.

Since we have the CI coefficients now as a function of $\xi_1$, we also have the 2RDM elements $\overline{\Gamma}^{D_{2h}}_{ij,kl}[^1B_{1g},\xi_1,\mat{n}]$, which are in principle those given in Ref.~\cite{GritsenkoWangKnowles2019}, except that there is a typo in the opposite-spin-block of the 2RDM for square H$_4$ Eq.~(22). The columns and rows of $2\bar{2}$ and $3\bar{3}$ should be empty, since these terms never occur in the wavefunction. The correct symmetry blocked 2RDM has been deferred to Appendix~\ref{sec:1B1g2RDM}, as it is rather unwieldy and not useful for the discussion at this point.

The ground state of the H$_2 + 2$H system belongs to the $^1A_g$ irrep for which the expansion becomes
\begin{align}
\Psi^{^1A_g} ={}& c^{ab}_{11}\abs[\big]{1a_g^2b_{2u}^2} + c^{aa}_{12}\abs[\big]{1a_g^22a_g^2} + 
c^{ab}_{12}\abs[\big]{1a_g^2b_{3u}^2} + {} \notag \\*
&c^{ab}_{21}\abs[\big]{2a_g^2b_{2u}^2} + c^{bb}_{12}\abs[\big]{b_{2u}^2b_{3u}^2} +
c^{ab}_{22}\abs[\big]{2a_g^2b_{3u}^2} + {} \notag \\*
&c^b_1\abs[\big]{b_{2u}^21a_g2a_g[\alpha\beta - \beta\alpha]} + {} \notag \\*
&c^b_2\abs[\big]{b_{3u}^21a_g2a_g[\alpha\beta - \beta\alpha]}  .
\end{align}
Because there are more terms, constructing a parametrisation becomes more tedious. Since there are 8 terms in the wavefunction and the 1RDM yields 5 non-trivial conditions (4 occupation numbers and 1 non-trivial off-diagonal element in the $a_g$ block), we expect that we need at least 3 parameters.

In order for this wavefunction to yield a diagonal 1RDM, we need to satisfy the following condition
\begin{equation}\label{eq:diagCondAgOrig}
c^b_1(c^{ab}_{11} + c^{ab}_{21}) + c^b_2(c^{ab}_{12} + c^{ab}_{22}) = 0 ,
\end{equation}
which can be rewritten as
\begin{multline}\label{eq:diagCondAg}
0 = (c^b_1 + c^b_2)(c^{ab}_{11} + c^{ab}_{22} + c^{ab}_{21} + c^{ab}_{12}) \\*
{} + (c^b_1 - c^b_2)(c^{ab}_{11} - c^{ab}_{22} + c^{ab}_{21} - c^{ab}_{12}) .
\end{multline}
The advantage of this form is that we can now eliminate $c^{ab}_{11} - c^{ab}_{22}$ and $c^{ab}_{21} - c^{ab}_{12}$ in favour of the other terms, by exploiting the following two conditions put by the occupation numbers on the coefficients
\begin{subequations}\label{eq:occCond}
\begin{align}
\Delta_+
&= \abs{c^{ab}_{11}}^2 - \abs{c^{ab}_{22}}^2 + \abs{c^b_1}^2 - \abs{c^b_2}^2 , \\
\Delta_-
&= \abs{c^{ab}_{12}}^2 - \abs{c^{ab}_{21}}^2 + \abs{c^b_2}^2 - \abs{c^b_1}^2 ,
\end{align}
\end{subequations}
where
\begin{equation}
\Delta_{\pm} = (n_{1a_g} - n_{2a_g} \pm n_{b_{2u}} \mp n_{b_{3u}})/4 .
\end{equation}
Introducing the following parametrisation for the coefficients, 
\begin{align}
\xi_1 &= c^{ab}_{11} + c^{ab}_{22} , &
\xi_2 &= c^{ab}_{12} + c^{ab}_{21} , &
\xi_3 &= c^b_1 - c^b_2 ,
\end{align}
the diagonality condition~\eqref{eq:diagCondAg} yields an explicit equation for $c^b_1 + c^b_2$
\begin{subequations}
\begin{align}\label{eq:zeta4}
\zeta_4 &= c^b_1 + c^b_2
= \frac{\xi_3}{\xi_1 + \xi_2}\frac{\xi_1\Delta_- - \xi_2\Delta_+}{\xi_1\xi_2 - \xi_3^2} .
\intertext{From the conditions~\eqref{eq:occCond} themselves we can extract}
\zeta_5 &= c^{ab}_{11} - c^{ab}_{22} = \bigl(\Delta_+ - \xi_3\zeta_4\bigr) / \xi_1 , \\
\zeta_6 &= c^{ab}_{12} - c^{ab}_{21} = \bigl(\Delta_- + \xi_3\zeta_4\bigr) / \xi_2 .
\intertext{The normalisation condition of the wavefunction, or equivalently the trace of the 1RDM, yields}
\label{eq:caa12pluscb12cond}
\zeta_7 &= \abs{c^{aa}_{12}}^2 + \abs{c^{bb}_{12}}^2 \\
&= 1 - \xi_3^2 - \zeta_4^2 - \thalf\bigl(\xi_1^2 + \xi_2^2 + \xi_5^2 + \xi_6^2\bigr) \notag
\intertext{and there is one additional independent linear combination of occupation numbers, which yields the relation}
\label{eq:caa12mincb12cond}
\zeta_8 &= \abs{c^{aa}_{12}}^2 - \abs{c^{bb}_{12}}^2 = \Delta_0 .
\end{align}
\end{subequations}
where
\begin{equation}\label{eq:Delta0def}
\Delta_0 = (n_{1a_g} + n_{2a_g} - n_{b_{2u}}  - n_{b_{3u}}) / 4
\end{equation}
The CI coefficients are obtained as
\begin{subequations}
\begin{align}
c^{ab}_{11} &= (\xi_1 + \zeta_5)/2 , &		c^{ab}_{22} &= (\xi_1 - \zeta_5)/2 , \\
c^{ab}_{12} &= (\xi_2 + \zeta_6)/2, &			c^{ab}_{21} &= (\xi_2 - \zeta_6)/2 , \\
c^b_1 &= (\xi_3 + \zeta_4)/2 , &	 			c^b_2 &= (\xi_3 - \zeta_4)/2 , \\
c^{aa}_{12} &= \sqrt{\zeta_7 + \zeta_8} , &	c^{bb}_{12} &= \xi_4\sqrt{\zeta_7 - \zeta_8} ,
\end{align}
\end{subequations}
where we needed to introduce one additional parameter $\xi_4 = \pm 1$ to handle the unknown relative phase factor.
Since the exact form of the 2RDM is not particularly enlightening, it is not presented here, but still reported in Appendix~\ref{sec:1Ag2RDM} for completeness.

\begin{table}[b]
\caption{The unique matrix elements of the spin-summed 2RDM for the determination of the relative sign of the $^1B_{1g}$ wavefunction coefficients.}
\label{tab:spinSummed2RDM}
\begin{ruledtabular}
\begin{tabular}{CCCCC}
i	&j			&k		&l			&\Gamma_{ij,kl}			\\
\midrule
1a_g	&1a_g		&2a_g	&2a_g		&4c_1c_2					\\
1a_g	&b _{2u}		&b _{2u}	&2a_g		&-6c_3(c_2 - c_1)			\\
1a_g	&b _{3u}		&b _{3u}	&2a_g		&\hphantom{-}6c_3(c_2 - c_1)	\\
\end{tabular}
\end{ruledtabular}
\end{table}

\section{Optimisation of the free parameters}
\label{sec:optimisation}

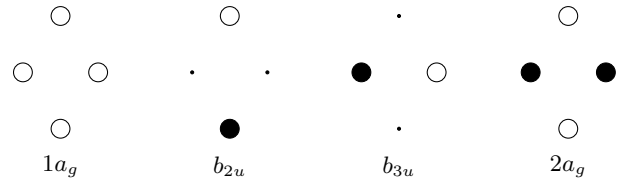
\begin{figure}[t]
\begin{center}
\begin{tikzpicture}
  \draw (0.5,0) circle (0.125);
  \draw (-0.5,0) circle (0.125);
  \draw (0,0.75) circle (0.125);
  \draw (0,-0.75) circle (0.125);
  \node at (0,-1.25) {$1a_g$};
  \begin{scope}[xshift=2.25cm]
    \filldraw (0.5,0) circle (0.02);
    \filldraw (-0.5,0) circle (0.02);
    \draw (0,0.75) circle (0.125);
    \filldraw (0,-0.75) circle (0.125);
    \node at (0,-1.25) {$b_{2u}$};
  \end{scope}
  \begin{scope}[xshift=4.5cm]
    \draw (0.5,0) circle (0.125);
    \filldraw (-0.5,0) circle (0.125);
    \filldraw (0,0.75) circle (0.02);
    \filldraw (0,-0.75) circle (0.02);
    \node at (0,-1.25) {$b_{3u}$};
  \end{scope}
  \begin{scope}[xshift=6.75cm]
    \filldraw (0.5,0) circle (0.125);
    \filldraw (-0.5,0) circle (0.125);
    \draw (0,0.75) circle (0.125);
    \draw (0,-0.75) circle (0.125);
    \node at (0,-1.25) {$2a_g$};
  \end{scope}
\end{tikzpicture}
\caption{Sketch of the NOs of the rhombic H$_4$ system which has $D_{2h}$ symmetry.}
\label{fig:NOs}
\end{center}
\end{figure}

Since the exact symmetry-restricted functionals derived in Sec.~\ref{sec:paramD2h} (and the more general $C_2$ functional in Ap.~\ref{ap:paramC2v}) still contain at least one parameter, they are not explicit 1RDM functionals. Though the precise values of the 2RDM matrix elements will vary for different systems, we can still make some general statements, especially for the $^1B_{1g}$ component of the $D_{2h}$ functional (Sec.~\ref{sec:paramD2h}), which only contains one parameter $\xi_1$. The dependence of the wavefunction coefficients is shown in Fig.~\ref{fig:B1gcoefs} for $n_{1a_g} - n_{2a_g} = 0.2$. One can see that the variable $\xi_1$ is able to generate all possible sign combinations of $c_1$ and $c_2$, so the first task will be to pinpoint the correct sign pattern. In Tab.~\ref{tab:spinSummed2RDM} the relevant spin-summed 2RDM matrix elements are reported. The first 2RDM element in Tab.~\ref{tab:spinSummed2RDM} corresponds to a positive two-electron integral, so $c_1c_2 < 0$ would be beneficial to reduce the repulsion. For the other two reported 2RDM elements in Tab.~\ref{tab:spinSummed2RDM} it is important to realise that their corresponding integrals have opposite signs, depending on the actual phase of the $2a_g$ NO: if the $2a_g$ NO is positive along the $b_{2u}$ orbital ($y$ direction) as sketched in Fig.~\ref{fig:NOs}, the integral $\braket{1a_g b _{2u}}{b _{2u} 2a_g} > 0$ and $\braket{1a_g b _{3u}}{b _{3u} 2a_g} < 0$ and vice versa. For either phase choice, this means that the signs of $c_1$ and $c_2$ should be opposite for both terms to have a maximally stabilising effect, which agrees with a stabilising first element in Tab.~\ref{tab:spinSummed2RDM}. For the phase choice of the $2a_g$ NO depicted in Fig.~\ref{fig:NOs}, this means $c_2 > 0$ and $c_1 < 0$, so
\begin{equation}
\xi_1 \in \left[-\half\sqrt{\Delta_{\pm}}, -\half\sqrt{2 - 2\sqrt{1 - \Delta_{\pm}^2}}\right] , 
\end{equation}
where $\Delta_{\pm} = n_{1a_g} - n_{2a_g}$, because $n_{b_{2u}} = n_{b_{3u}} = 1$ is necessarily in this sector, cf.\ Sec.~\ref{sec:paramD2h}. For the opposite phase choice, the relevant interval for $\xi_1$ would need to be reflected w.r.t.\ the origin.

This immediately signals a problem when trying to develop an explicit proper approximate 1RDM functional in terms of the NOs. If we would completely fix the sign of the expansion coefficients, e.g.\ $c_1 > 0$ and $c_2 < 0$, i.e.\ $\xi_1 > 0$, the functional would become phase dependent and not be a pure 1RDM functional anymore.

Let us investigate the size of the error by considering two rhombic H$_4$ systems from Ref.~\cite{GritsenkoWangKnowles2019} with an apex of 90\degree\ ($R = 2.0$ Å), i.e.\ a square, and with an apex of 120\degree\ ($R = 1.972\,665\,297\,958\,2$ Å)~\footnote{The optimal distance has been determined with a few additional digits, so that the natural occupation numbers do not differ by more than $10^{-13}$.}.
The calculations have been performed with the help of the full CI module of \textsc{pyscf}~\cite{PYSCF, Sun2015}, again in the STO-3G basis~\cite{STO-nG_H-Ne}.
We see from the 2RDM elements in their respective NO representations reported in Tab.~\ref{tab:SpinDifference} that an incorrect sign leads to large deviations in the 2RDM elements. If the correct sign is used, the maximum deviation in the 2RDM elements is in the order of $10^{-4}$. Using the 2RDM elements of the $90\degree$ rhombus for the calculation of the total energy of the $120\degree$ rhombus and vice versa leads to an energy difference in the order of micro Hartrees: 1.280\,$\mu$H and 1.322\,$\mu$H respectively. However, using the 2RDM elements with the incorrect sign leads to errors of almost 1 Hartree: 0.735\,H and 0.761\,H respectively.

\begin{table}[b]%
\centering%
\caption{Comparison of all differences larger than $10^{-4}$ 2RDM matrix elements in their respective NO basis, where $k=2,3$. The reported difference is for the most optimal phase choice of the $a_g$ NOs (upper sign).}
\label{tab:SpinDifference}
{\footnotesize Largest differences in the like-spin 2RDM elements $\Gamma^{\sigma\sigma}_{ij,kl}$.}
\begin{ruledtabular}
\begin{tabular}{CCCCd{6}d{6}d{6}}
i	&j		&k		&l		&\multicolumn{1}{c}{90\degree}
		&\multicolumn{1}{c}{120\degree}	&\multicolumn{1}{c}{difference} \\
\midrule
1a_g		&b_{2u}	&b_{2u}	&2a_g	&\mp0.106392	&-0.106260	&-0.000132	\\
1a_g		&b_{3u}	&b_{3u}	&2a_g	&\pm0.106392	&0.106260	&0.000132	\\
\midrule
1a_g		&2a_g	&1a_g	&2a_g	&0.065563	&0.065317	&0.000247	\\
b_{2u}	&b_{3u}	&b_{2u}	&b_{3u}	&0.065563	&0.065317	&0.000247	\\
1a_g		&b_{ku}	&1a_g	&b_{ku}	&0.356722	&0.356845	&-0.000123	\\
2a_g		&b_{ku}	&2a_g	&b_{ku}	&0.077715	&0.077838	&-0.000123	\\
\end{tabular}
\end{ruledtabular}
\vspace{\baselineskip}
{\footnotesize Largest differences in the opposite-spin 2RDM elements $\Gamma^{\alpha\beta}_{ij,kl}$.}
\begin{ruledtabular}
\begin{tabular}{CCCCd{6}d{6}d{6}}
i	&j		&k		&l		&\multicolumn{1}{c}{90\degree}
		&\multicolumn{1}{c}{120\degree}	&\multicolumn{1}{c}{difference}	\\
1a_g		&\bar{b}_{2u}	&b_{2u}	&2\bar{a}_g	&\mp0.212784	&-0.212520	&-0.000265	\\
1a_g		&\bar{b}_{3u}	&b_{3u}	&2\bar{a}_g	&\pm0.212784	&0.212520	&0.000265	\\
\midrule
1a_g		&1\bar{a}_g	&2a_g	&2\bar{a}_g	&-0.288932	&-0.289447	&0.000514	\\
1a_g		&1\bar{a}_g	&1a_g	&1\bar{a}_g	&0.680662	&0.681032	&-0.000370	\\
2a_g		&2\bar{a}_g	&2a_g	&2\bar{a}_g	&0.122648	&0.123018	&-0.000370	\\
b_{2u}	&\bar{b}_{3u}	&b_{2u}	&\bar{b}_{3u}	&0.434437	&0.434683	&-0.000247	\\
1a_g		&2\bar{a}_g	&1a_g	&2\bar{a}_g	&0.032782	&0.032658	&0.000123	\\
1a_g		&\bar{b}_{ku}	&1a_g	&\bar{b}_{ku}	&0.422285	&0.422162	&0.000123	\\
2a_g		&\bar{b}_{ku}	&2a_g	&\bar{b}_{ku}	&0.143278	&0.143155	&0.000123	\\
\end{tabular}
\end{ruledtabular} 
\end{table}

\begin{table}[b]
\caption{The $2a_g$ NO represented in the AO basis (STO-3G) for different apices of H$_4$ in a rhombic arrangement with adjusted }
\label{tab:rhombicNOs}
\begin{ruledtabular}
\begin{tabular}{cd{3}d{3}d{3}}
		&\multicolumn{1}{c}{60\degree}	&\multicolumn{1}{c}{90\degree}	&\multicolumn{1}{c}{120\degree}	\\
\midrule
H$_1$	&0.576					&\mp0.563				&-0.532	\\
H$_2$	&0.576					&\mp0.563				&-0.532	\\
H$_3$	&-0.532					&\pm0.563				&0.576	\\
H$_4$	&-0.532					&\pm0.563				&0.576	\\
\end{tabular}
\end{ruledtabular}
\end{table}

One could hope that there would be some kind of universal choice for the NOs phase, e.g.\ choosing the sign of the largest coefficient in the NO to be positive. However, such an approach appears not to provide a solution. In Tab.~\ref{tab:rhombicNOs} the coefficients of the $2a_g$ NO of a rhombic H$_4$ system with an apex of $60\degree$ and with an apex of $120\degree$ are reported. Though these systems are completely equivalent, so basically have identical 2RDM matrix elements in their NO representation, the convention of choosing the coefficient with the largest amplitude to be positive causes the $2a_g$ NO to have an opposite sign in both systems. Hence, we obtain large differences in the 2RDM elements and corresponding total energies. which are of the same order as in the previous example. The proposed sign convention is even more problematic in the case of square H$_4$ (90\degree): none of the coefficients is larger in magnitude than the others, so the convention is indecisive.

The problem is aggravated if the more general $C_2$ functional is used (see Appendix~\ref{ap:paramC2v} for a full parametrisation). Using this lower symmetry, the $b_{2u}$ and $b_{3u}$ orbitals are not separated by symmetry anymore and are allowed to form any unitary combination, because they have the same natural occupation number ($n_{b_{2u}} = n_{b_{3u}} = 1$). Because the shape of these mixtures can vary arbitrarily between different systems, their correspondence is lost and can lead to deviations in the 2RDM matrix elements of similar magnitude as in the case of the undetermined phase of the $2a_g$ NO.

Therefore, a proper approximate 1RDM functional cannot be an explicit functional in general in NO representation, if it is supposed to be invariant under unitary transformations of degenerate NOs%
~\footnote{Note that this includes phase invariance, since any eigenfunction can be considered to be degenerate with itself.}.
However, one can take different viewpoints of this situation. The essence of 1RDM functional theory is to split the minimisation of the energy in terms of the many-body wavefunction in terms of the 1RDM and the remaining degrees of freedom. Using a 1RDM functional which does not respect unitary invariance of degenerate NOs (and phase invariance) in this scheme, simply means that some of the remaining degrees of freedom will be combined with the 1RDM optimisation. Taking this view already at the theoretical level means that one should rather speak of an NO functional theory, which seems to be in line with the view of Piris~\cite{Piris2007} and also the idea of phase including NOs proposed in the time-dependent setting~\cite{GiesbertzGritsenkoBaerends2010a, GiesbertzGritsenkoBaerends2010b, GiesbertzGritsenkoBaerends2012b, MeerGritsenkoGiesbertz2013, GiesbertzGritsenkoBaerends2014}. So indeed, this would mean the extension of 1RDM functional theory that Wang and Knowles were referring to in Ref.~\cite{WangKnowles2015}. The downside of this view at a theoretical level is that a one-to-one relation with the non-local potential would definitely be out of the question on dimensional grounds, also at elevated temperatures~\cite{GiesbertzRuggenthaler2019}.

An other option is to impose this viewpoint at the implementation level. One still accepts that one needs to deal with an implicit 1RDM functional, but that the optimisation of the parameters related to the unitary invariance of degenerate NOs is shifted to the optimisation of NOs to obtain a more practical implementation of the implicit functional. For the minimisation of the total energy all parameters should be optimised and it is irrelevant for the end result how we group the variables together.
However, for the calculation of response properties one should carefully scrutinise how the invariance of the functional should be taken into account.

To finalise this section, let us also briefly discuss the $^1A_g$ component of the $D_{2h}$ functional.
It might seem to be a contradiction that we still have 4 free parameters for the H$_2 + 2$H system, since Gritsenko et al.\ in Ref.~\cite{GritsenkoWangKnowles2019} did not find additional parameters. The difference is that we only used symmetry to build the restricted functional. However, if we also use the special property of the H$_2 + 2$H system that the Coulomb integrals between the fragments are zero in this limit, the minimisation in~\eqref{eq:symAdapIrrep} can be executed explicitly (see Appendix~\ref{sec:explicitMin}) and we recover the result by Gritsenko et al.~\cite{GritsenkoWangKnowles2019}: a one-to-one relation between the 2RDM elements and the occupation numbers.

Such additional assumptions are often made to develop practical approximations, geared towards the physical situation one is interested in. We actually made such an assumption ($c_4 = 0$) for $\overline{\Gamma}^{D_{2h}}_{ij,kl}[^1B_{1g},\xi_1,\mat{n}]$ to simplify the final expression. A similar assumption is well known for the singlet two-electron case, where the phase $\e^{\im\xi_k}$ is taken to be positive for the highest occupied NO and negative for all other NOs~\cite{GoedeckerUmrigar2000}. However, this additional assumption reduces the validity of the two-electron functional as it is not exact anymore for all singlet two-electron cases~\cite{Cioslowski2018}. Though the covalent bonding is still correctly described, the Van der Waals interactions are missing~\cite{CioslowskiPernal2006, MentelShengGritsenko2012, GiesbertzLeeuwen2013b, CioslowskiPratnicki2019}.

\section{Conclusion}
\label{sec:conclusion}

It has been stressed that the many-to-one relation of the map $\mat{\Gamma} \mapsto \mat{n}$ in~\eqref{eq:2RDMtoOccMap} is responsible for the possibility that identical occupation numbers can correspond to different 2RDMs.
The constrained-search functional~\eqref{eq:NOconstrSearch}, or equivalently~\eqref{eq:freeSearch}, determines these remaining degrees of freedom by minimising over the contraction of the 2RDM with the two-electron integrals in the NO basis
Thus, different NOs lead to different two-electron integrals, which lead to a different minimum within the constrained-search functional~\eqref{eq:NOconstrSearch} and hence, to different 2RDMs. No symmetry is needed to explain this property.

An explicit construction of a constrained-search functional can be achieved by identifying the remaining degrees of freedom in the wavefunction~\eqref{eq:LevySearch} or even density-matrix operator~\eqref{eq:ValoneSearch}. This is most easily achieved for the pure state case and the general procedure has been outlined at the end of Sec.~\ref{sec:symAdapt}. This procedure has been exemplified by an explicit parametrisation of the 2RDM matrix elements for the two-electron functional~\eqref{eq:exactTwoRDM} and the construction of several irreps of symmetry-restricted functionals for four electrons in four $s$-orbitals (Sec.~\ref{sec:paramD2h} and Ap.~\ref{ap:paramC2v}).
The main reason to impose symmetry restrictions on the (validity of) the functional is that the constrained search can be broken down over the irreps of the symmetry group~\eqref{eq:symRestrFunc} and the variational freedom within each irrep of the group is significantly reduced. This reduced variational freedom allows one to more easily find an explicit parametrisation and makes the parametrisation also more transparent than a parametrisation of the original functional without symmetry restrictions. The disadvantage is, of course, that the symmetry-restricted functional is only valid for systems exhibiting the presumed symmetry. However, such exact parametrisations could serve as a guide in the construction of more general approximate 1RDM functionals, by revealing the necessary structures to incorporate additional flexibility.

One can wonder whether we should invest in building approximate symmetry-restricted 1RDM functionals, especially in light of the fact that explicit symmetry dependence reduces the validity of the functional to only a symmetry class of potentials / systems. Moreover, since 1RDM functional theory actually aims to be practical for medium sized systems that often do not display any symmetry, a 1RDM functional relying on symmetry would be of limited use. An exception for finite systems are the spin states, since usually our molecular Hamiltonian is spin-independent. Approximations towards this direction can be found in Refs~\cite{PirisLopezRuiperez2011, Piris2019}, but have recently been criticised for violating important constraints~\cite{Cioslowski2020}.
Being able to access the different spin states is very useful in the study of transition-metal compounds. For infinite models of the bulk of crystals, translational symmetry can be exploited~\cite{SchillingSchilling2019}.

It is worthwhile to search also for other possibilities to reduce the validity of the functional, because, inherited from the many-body wavefunction / density-matrix operator, the exact functional contains an exponentially growing number of parameters $\mat{\xi}$. This makes an explicit construction of the functional infeasible and even undesirable for practical purposes. By reducing the number of potentials for which the functional needs to be valid, the flexibility of the wavefunction / density-matrix operator in the constrained-search functional can be reduced, so fewer parameters $\mat{\xi}$ need to be included in the functional. Another option to reduce the validity of the functional is to exploit that we are typically only interested in a very limited class of external potentials, e.g.\ Coulomb potentials.

The explicitly parametrised irrep components of the $D_{2h}$ symmetry-restricted functional for four electrons in four $s$-orbitals in Sec.~\ref{sec:paramD2h} are relevant for the square H$_4$ and H$_2 + 2$H systems considered by Wang and Knowles~\cite{WangKnowles2015}. They use these H$_4$ systems to give an explicit example of two systems with identical occupation numbers, but different 2RDM elements. This parametrisation is general enough to be also valid for the sequence of rhombi studied later in joint work with Gritsenko~\cite{GritsenkoWangKnowles2019}.
In that work, the authors showed that the differences in the 2RDM eigenvalues for two H$_4$ geometries tend to be small if the underlying wavefunctions have the same symmetry, so out of scope in the construction of \emph{approximate} functionals, which they refer to as a \emph{practical functional}. Only when the underlying states belong to a different irrep were appreciable differences in the 2RDM eigenvalues found.
This situation sounds like a state crossing, but around an apex of $125\degree$, no length $R$ could be found anymore to retain identical occupation numbers, so the crossing point cannot be reached for the rhombi.
This finding is of course due to the completely different character of the underlying wavefunctions which could not be differentiated by the occupation numbers. This is the essential aspect which pertains to the exact universal functional~\eqref{eq:LevySearch} or~\eqref{eq:ValoneSearch} which is symmetry independent of course.
One therefore expects that this situation could also occur for states with identical symmetry but still different character (avoided crossings). Although, that the occupation numbers are identical is a quite stringent condition. So it seems an unlikely situation in practice, especially when a more realistic number of orbitals is taken into account. However, perhaps for a model system like H$_4$ or He$_2$H$_2^{2+}$ in a minimal basis, two geometries can be found with identical occupation numbers and electronic states not differentiable by symmetry, but with sizeable differences in the 2RDM elements.

In this paper, I have also shown that the chosen phase of the NOs can be important for the sign of the 2RDM matrix elements, so actually prohibits proper 1RDM functionals in which the 2RDM matrix elements only depend explicitly on the occupation numbers, $\Gamma_{ij,kl}(n)$. Of course, the Hartree and exchange 2RDM elements, $\Gamma_{ij,ij}$ and $\Gamma_{ij,ji}$ respectively are invariant with respect to the NO phase, but is has become clear that other 2RDM elements are important if we want to construct more reliable functionals~\cite{MentelShengGritsenko2012, MentelMeerGritsenko2014, Piris2017, MeerGritsenkoBaerends2018, Cioslowski2020}. A proper 1RDM functional needs to be invariant with respect to the phase of the NOs, since the phase is undetermined, as the NOs are eigenfunctions of a hermitian operator. So if one aims to build a proper approximate 1RDM functional beyond Hartree and exchange ($JK$-functional), one can not do without an additional set of parameters $\mat{\xi}$ to make the functional invariant with respect to the phase of the NOs. This is quite easy to achieve by placing $\e^{\im(\xi_k + \xi_l - \xi_i - \xi_j)}$ in front of each 2RDM matrix element, c.f.\ the exact two-electron functional~\eqref{eq:exactTwoRDM}.
Of course, in a practical implementation in which the occupation numbers and NOs are optimised directly, one can leave this additional degree of freedom out of the functional and let the NO phase take care of this part of the functional.
Alternatively, one could already at the theoretical level say that we are working instead with an NO functional theory, so the functionals are allowed to depend on the (relative) phases of the NOs. This line of thought is followed by Piris~\cite{Piris2007, Piris2014, Piris2017} and is also in the ``phase including natural orbital'' approach to resolve the many pathological issues encountered while formulating an adiabatic approximate in time-dependent 1RDM functional theory~\cite{PhD-Giesbertz2010, GiesbertzGritsenkoBaerends2010a, GiesbertzGritsenkoBaerends2010b, GiesbertzGritsenkoBaerends2012b, MeerGritsenkoGiesbertz2013, GiesbertzGritsenkoBaerends2014}.

\begin{acknowledgments}
The author would like to thank prof.\ dr.\ E.J. Baerends and dr.\ O.V. Gritsenko for useful comments and discussions.
The author acknowledges funding by Stichting voor Fundamenteel Onderzoek der Materie FOM Projectruimte [project 15PR3232] and European Research Council under H2020/ERC Consolidator Grant ``corr-DFT'' (Grant No.\ 648932)
\end{acknowledgments}

\appendix

\section{2RDM of the $^1B_{1g}$ state}
\label{sec:1B1g2RDM}

We now discuss the 2RDM of the $^1B_{1g}$ state under the assumption $n_{1a_g} \neq n_{2a_g}$. Due to the symmetry, the 2RDM becomes symmetry blocked. Abbreviating the orbitals as $1_a = 1a_g$, $2_a = 2a_g$, $1_b = b_{2u}$ and $2_b = b_{3u}$, the like-spin blocks become
\begin{gather}%
\kbordermatrix{
A_g^{\uparrow\uparrow}	&1_a 2_a	\\
1_a 2_a			&4c_3^2
} , \\
\kbordermatrix{
B_{1g}^{\uparrow\uparrow}	&1_b2_b	\\
1_b2_b				&4c_3^2
} , \\
\kbordermatrix{
B_{2u}^{\uparrow\uparrow}	&1_a 1_b	 	&2_a 1_b		\\
1_a 1_b				&c_1^2 + c_3^2	&h				\\
2_a 1_b				&h				&c_2^2 + c_3^2
} , \\
\kbordermatrix{
B_{3u}^{\uparrow\uparrow}	&1_a 2_b	 	&2_a 2_b		\\
1_a 2_b				&c_1^2 + c_3^2	&-h				\\
2_a 2_b				&-h				&c_2^2 + c_3^2
} ,
\end{gather}
and the opposite-spin blocks become
\begin{gather}
\kbordermatrix{
A_g^{\uparrow\downarrow}	&1_a \bar{1}_a	&1_a \bar{2}_a	&2_a \bar{1}_a	&2_a \bar{2}_a
		&1_b\bar{1}_b		&2_b\bar{2}_b	\\
1_a \bar{1}_a			&2c_1^2			&0				&0				&2c_1c_2	&0	&0	\\
1_a \bar{2}_a			&0				&2e				&-2e				&0		&0	&0	\\
2_a \bar{1}_a			&0				&-2e				&2e				&0		&0	&0	\\
2_a \bar{2}_a			&2c_1c_2			&0				&0				&2c_2^2	&0	&0	\\
1_b\bar{1}_b			&0				&0				&0				&0		&0	&0	\\
2_b\bar{2}_b			&0				&0				&0				&0		&0	&0
} , \\
\kbordermatrix{
B_{1g}^{\uparrow\downarrow}	&1_b \bar{2}_b	&2_b \bar{1}_b	 \\
1_b \bar{2}_b			&d + 2e			&d - 2e			\\
2_b \bar{1}_b			&d - 2e			&d + 2e
} , \\
\kbordermatrix{
B_{2u}^{\uparrow\downarrow}	&1_a \bar{1}_b	&2_a \bar{1}_b	&1_b\bar{1}_a	&1_b\bar{2}_a	 \\
1_a \bar{1}_b			&a				&-h				&4e				&-2h				\\
2_a \bar{1}_b			&-h				&b				&-2h				&4e				\\
1_b\bar{1}_a			&4e				&-2h				&a				&-h				\\
1_b\bar{2}_a			&-2h				&4e				&-h				&b
} , \\
\kbordermatrix{
B_{3u}^{\uparrow\downarrow}	&1_a \bar{2}_b	&2_a \bar{2}_b	&2_b\bar{1}_a	&2_b\bar{2}_a	 \\
1_a \bar{2}_b			&a				&h				&4e				&2h				\\
2_a \bar{2}_b			&h				&b				&2h				&4e				\\
2_b\bar{1}_a			&4e				&2h				&a				&h				\\
2_b\bar{2}_a			&2h				&4e				&h				&b
} ,
\end{gather}
where $a = c_1^2 + 5c_3^2$, $b = c_2^2 + 5c_3^2$, $d = c_1^2 + c_2^2$, $e = c_3^2$ and $h = c_3(c_2 - c_1)$.

\section{2RDM of the $^1A_{g}$ state}
\label{sec:1Ag2RDM}
Abbreviating the orbitals as $1_a = 1a_g$, $2_a = 2a_g$, $1_b = b_{2u}$ and $2_b = b_{3u}$, the like-spin blocks of the 2RDM of the $^1A_g$ state become
\begin{gather}
\kbordermatrix{
A_g^{\uparrow\uparrow}	&1_a 2_a	\\
1_a 2_a				&\abs{c^{aa}_{12}}^2
} , \\
\kbordermatrix{
B_{1g}^{\uparrow\uparrow}	&1_b 2_b	\\
1_b 2_b					&\abs{c^{bb}_{12}}^2
} , \\
\kbordermatrix{
B_{2u}^{\uparrow\uparrow}	&1_a 1_b		&2_a 1_b	\\
1_a 1_b					&s_{11}		&t_1 		\\
2_a 1_b					&t_1			&s_{21}
} , \\
\kbordermatrix{
B_{3u}^{\uparrow\uparrow}	&1_a 2_b		&2_a 2_b	\\
1_a 2_b					&s_{12}		&t_2		\\
2_a 2_b					&t_2			&s_{22}
} ,
\end{gather}
where
\begin{subequations}\label{eq:stDefs}
\begin{align}
s_{ij}	&= \abs{c^{ab}_{ij}}^2 + \abs{c^b_j}^2 ,	\\
t_j	&= c^b_j(c^{ab}_{1j} + c^{ab}_{2j}) .
\end{align}
\end{subequations}
The opposite-spin blocks become
\begin{equation}
\kbordermatrix{
A_g^{\uparrow\downarrow}	&1_a \bar{1}_a	&1_a \bar{2}_a	&2_a \bar{1}_a
	&2_a \bar{2}_a	&1_b \bar{1}_b	&2_b \bar{2}_b	\\
1_a \bar{1}_a	&d^a_1	&p_1		&p_1		&m		&l_{11}	&l_{12}	\\
1_a \bar{2}_a	&p_1		&d^a_m	&n		&p_2		&k_1		&k_2		\\
2_a \bar{1}_a	&p_1		&n		&d^a_m	&p_2		&k_1		&k_2		\\
2_a \bar{2}_a	&m		&p_2		&p_2		&d^a_2	&l_{21}	&l_{22}	\\
1_b \bar{1}_b	&l_{11}	&k_1		&k_1		&l_{21}	&d^b_1	&q		\\
2_b \bar{2}_b	&l_{12}	&k_2		&k_2		&l_{22}	&q		&d^b_2
} ,
\end{equation}
where
\begin{subequations}
\begin{align}
d^a_i	&= \abs{c^{aa}_{12}}^2 + \abs{c^{ab}_{i1}}^2 + \abs{c^{ab}_{i2}}^2 , \\
d^a_m	&= \abs{c^{aa}_{12}}^2 + \abs{c^b_1}^2 + \abs{c^b_2}^2 , \\
d^b_i	&= \abs{c^{bb}_{12}}^2 + \abs{c^{ab}_{1i}}^2 + \abs{c^{ab}_{2i}}^2 + 2\abs{c^b_i}^2 , \\
p_i		&= c^{ab}_{i1}c^b_1 + c^{ab}_{i2}c^b_2 , \\
k_i		&= c^{bb}_{12}c^b_{3-i} - c^{aa}_{12}c^b_i , \\
l_{ij}		&= c^{ab}_{(3-i)j}c^{aa}_{12} + c^{ab}_{i(3-j)}c^{bb}_{12} , \\
m		&= c^{ab}_{11}c^{ab}_{21} + c^{ab}_{12}c^{ab}_{22} , \\
n		&= \abs{c^b_1}^2 + \abs{c^b_2}^2 , \\
q		&= c^{ab}_{12}c^{ab}_{11} + c^{ab}_{21}c^{ab}_{22} + c^b_1c^b_2
\end{align}
\end{subequations}
The other blocks are
\begin{gather}
\kbordermatrix{
B_{1g}^{\uparrow\downarrow}	&1_b \bar{2}_b			&2_b \bar{1}_b	\\
1_b \bar{2}_b				&\abs{c^{bb}_{12}}^2		&0			\\
2_b \bar{1}_b				&0					&\abs{c^{bb}_{12}}^2	
} , \\
\kbordermatrix{
B_{2u}^{\uparrow\downarrow} 	&1_a \bar{1}_b	&2_a \bar{1}_b	&1_b \bar{1}_a	&1_b \bar{2}_a	\\
1_a \bar{1}_b				&s_{11}		&t_1			&0			&0			\\
2_a \bar{1}_b				&t_1			&s_{21}		&0			&0			\\
1_b \bar{1}_a				&0			&0			&s_{11}		&t_1			\\
1_b \bar{2}_a				&0			&0			&t_1			&s_{21}
} , \\
\kbordermatrix{
B_{3u}^{\uparrow\downarrow}	&1_a \bar{2}_b	&2_a \bar{2}_b	&2_b \bar{1}_a	&2_b \bar{2}_a	\\
1_a \bar{2}_b				&s_{12}		&t_2			&0			&0			\\
2_a \bar{2}_b				&t_2			&s_{22}		&0			&0			\\
2_b \bar{1}_a				&0			&0			&s_{12}		&t_2			\\
2_b \bar{2}_a				&0			&0			&t_2			&s_{22}
}
\end{gather}
where $s_{ij}$ and $t_i$ were defined in~\eqref{eq:stDefs}.

\section{Explicit minimisation for the H$_2 + 2$H system}
\label{sec:explicitMin}
The advantage of this system is that most two-electron integrals are zero due to the distances. The only non-vanishing two-electron integrals are
\begin{equation}
\begin{split}
&\braket{1_a1_a}{1_a1_a} , \eqspace \braket{1_a1_b}{1_a1_b} ,  \\
&\braket{1_a1_b}{1_b1_a} , \eqspace \braket{1_b1_b}{1_b1_b} ,  \\
&\braket{2_a2_a}{2_a2_a} =
\braket{2_a2_b}{2_a2_b} = w_H \\
&\braket{2_a2_b}{2_b2_a} =
\braket{2_b2_b}{2_b2_b} = w_H
\end{split}
\end{equation}
This means that all blocks with $1_x2_y$ pairs ($x,y \in \{a,b\}$) do not contribute and the interaction energy expression becomes
\begin{multline}
W = s_{11}\bigl(3\braket{1_a1_b}{1_a1_b}  - \braket{1_a1_b}{1_b1_a}\bigr) + {} \\
d^a_1\braket{1_a1_a}{1_a1_a} + \thalf n_{1b}\braket{1_b1_b}{1_b1_b} + 2l_{11}\braket{1_a1_b}{1_b1_a} + {} \\
\bigl(d^a_2+ 2l_{22} + \thalf n_{2b} + 2s_{22}\bigr)w_H .
\end{multline}
All terms only contain squares of the coefficients, so are always positive, except the $l_{ii}$ terms which read
\begin{align}
l_{11} &= c^{ab}_{21}c^{aa}_{12} + c^{ab}_{12}c^{bb}_{12} , &
l_{22} &= c^{ab}_{12}c^{aa}_{12} + c^{ab}_{21}c^{bb}_{12} .
\end{align}
So if we choose
\begin{align}\label{eq:dissoPhaseChoice}
c^{aa}_{12} &> 0 &	
&\Rightarrow&
&c^{ab}_{12} < 0, c^{ab}_{21} < 0, c^{bb}_{12} > 0 ,
\end{align}
we minimise the interaction energy. Now it is useful to write out the positively contributing terms explicitly
\begin{align}
W
&= \bigl(\abs{c^{ab}_{11}}^2 + \abs{c^b_1}^2\bigr)
\bigl(3\braket{1_a1_b}{1_a1_b}  - \braket{1_a1_b}{1_b1_a}\bigr) + {} \notag \\
&\eqspace
\bigl(\abs{c^{aa}_{12}}^2 + \abs{c^{ab}_{11}}^2 + \abs{c^{ab}_{12}}^2\bigr)\braket{1_a1_a}{1_a1_a} + {} \notag \\
&\eqspace
\thalf n_{1b}\braket{1_b1_b}{1_b1_b} + 2l_{11}\braket{1_a1_b}{1_b1_a} + {} \\
&\eqspace
\bigl(\abs{c^{aa}_{12}}^2 + \abs{c^{ab}_{21}}^2 + 3\abs{c^{ab}_{22}}^2 + {} \notag \\
&\eqspace\hphantom{\bigl(}
2\abs{c^b_2}^2 + 2l_{22} + \thalf n_{2b}\bigr)w_H . \notag
\end{align}
We can minimise this expression if we can minimise $\abs{c^{ab}_{11}}^2$, $\abs{c^{ab}_{22}}^2$, $\abs{c^b_1}^2$ and $\abs{c^b_2}^2$, since they dominate over $\abs{c^{aa}_{12}}^2$, $\abs{c^{ab}_{12}}^2$, $\abs{c^{ab}_{21}}^2$ and $\abs{c^{bb}_{12}}^2$ terms because the latter also make negative contributions. The difference $\abs{c^{ab}_{11}}^2 - \abs{c^{ab}_{22}}^2$ is fixed by the occupation numbers~\eqref{eq:caa12mincb12cond}, so the best we can do is
\begin{subequations}
\begin{align}
c^{aa}_{12} &= \sqrt{\max(0,\Delta_0)} , \\
c^{bb}_{12} &= \sqrt{\max(0,-\Delta_0)} ,
\end{align}
\end{subequations}
where $\Delta_0$ was defined in~\eqref{eq:Delta0def}. Note that the phase of these coefficients does not matter, since they do not appear in any cross term in the interaction energy expression.

Now the parametrisation becomes very useful, since now we can vary over $\xi_1$, $\xi_2$ and $\xi_3$ without worrying about the constraints. Setting $\xi_3 = 0$, implies that $\zeta_4=0$~\eqref{eq:zeta4}, so we can even achieve $\abs{c^b_1}^2 = \abs{c^b_2}^2 = 0$. The expression for the interaction then reduces to
\begin{align}
W
&= \abs{c^{ab}_{11}}^2\bigl(3\braket{1_a1_b}{1_a1_b}  - \braket{1_a1_b}{1_b1_a}\bigr) + {} \notag \\
&\eqspace
\thalf n_{1a}\braket{1_a1_a}{1_a1_a} + \thalf n_{1b}\braket{1_b1_b}{1_b1_b} + {} \notag \\
&\eqspace
\bigl(2\abs{c^{ab}_{22}}^2+ \thalf (n_{2a} + n_{2b})\bigr)w_H + {} \\
&\eqspace
2l_{11}\braket{1_a1_b}{1_b1_a} + 2l_{22} w_H . \notag
\end{align}
All terms on the first three lines are now fixed and only the last line needs to be minimised. The last line can be rewritten as
\begin{equation}\label{eq:redInter}
\tilde{W}(\xi_1,\xi_2) = \xi_1\xi_2 w_+ + \frac{\Delta_+}{\xi_1}\frac{\Delta_-}{\xi_2}w_- ,
\end{equation}
where $w_{\pm} = w_H \pm \braket{1_a1_b}{1_b1_a}$ and $2w_H \geq w_+ \geq w_- \geq 0$.
The remaining variables are constrained by the normalisation~\eqref{eq:caa12pluscb12cond} as
\begin{equation}\label{eq:redConstr}
\half\biggl(\xi_1^2 + \frac{\Delta_+^2}{\xi_1^2} + \xi_2^2 + \frac{\Delta_-^2}{\xi_2^2}\biggr) = 1 - \abs{\Delta_0} .
\end{equation}
We can solve this equation to get an expression for $\xi_1$ in terms of $\xi_2$. Due to the phase convention~\eqref{eq:dissoPhaseChoice}, we should choose the positive root and since $w_+ > w_-$, we should choose the highest root for $\xi_1$, i.e.
\begin{multline}\label{eq:xi1inxi2}
\xi_1 = \frac{-1}{\sqrt{2}\xi_2}\Biggl[2(1 - \abs{\Delta_0})\xi_2^2 - \xi_2^4 - \Delta_-^2 + {} \\
\sqrt{\bigl(2(1 - \abs{\Delta_0})\xi_2^2 - \xi_2^4 - \Delta_-^2\bigr)^2 - 4\Delta_+^2\xi_2^4}\Biggr]^{\mathrlap{1/2}} .
\end{multline}
Now we insert this expression for $\xi_1$ back into~\eqref{eq:redInter} and find its stationary points
\begin{equation}
0 = \frac{\ud \tilde{W}}{\ud \xi_2} 
= \frac{\ud\tilde{W}}{\ud [\cdot]^{1/2}}\frac{\ud [\cdot]^{1/2}}{\ud\xi_2^2}\frac{\ud \xi_2^2}{\ud \xi_2} ,
\end{equation}
where $[\cdot]$ is the part in square brackets in~\eqref{eq:xi1inxi2}. Since the solution $\xi_2 = 0$ is not suitable, one of the other derivatives needs to vanish. Let us first consider vanishing of the first derivative on the r.h.s. 
\begin{equation}
0 = \frac{\ud\tilde{W}}{\ud [\cdot]^{1/2}}
= \frac{-1}{\sqrt{2}}\left[w_+ - \frac{2\Delta_+\Delta_-w_-}{[\cdot]}\right] .
\end{equation}
This equation is effectively a quadratic equation in $\xi_2^2$, so can be solved to yield
\begin{align}
\xi_2^2 &= \biggl(\frac{\Delta_+}{\Delta_-}\frac{w_+}{w_-} + 1\biggr)^{-1}\biggl[
1 - \abs{\Delta_0} \pm {} \\
&\eqspace
\sqrt{(1 - \abs{\Delta_0})^2 - (\Delta_+\tfrac{w_+}{w_-} + \Delta_-)(\Delta_+\tfrac{w_-}{w_+} + \Delta_-)}\biggr] , \notag
\end{align}
where we should choose the largest root, since the first term in~\eqref{eq:redInter} is dominant.

The other option is that the middle derivative on the r.h.s.\ of~\eqref{eq:xi1inxi2} vanishes, \( \ud [\cdot]^{1/2} / \ud\xi_2^2 = 0 \), which yields
\begin{align}
0 &= \frac{\ud [\cdot]^{1/2} }{ \ud\xi_2^2}
= \frac{1}{[\cdot]^{1/2}}\Biggl(1 - \abs{\Delta_0} - \xi_2^2 + {} \\*
&\frac{\bigl(2(1 - \abs{\Delta_0})\xi_2^2 - \xi_2^4 - \Delta_-^2\bigr)\bigl(1 - \abs{\Delta_0} - \xi_2^2\bigr)
 - 2\Delta_+^2\xi_2^2}{\sqrt{[2(1 - \abs{\Delta_0})\xi_2^2 - \xi_2^4 - \Delta_-^2]^2 - 4\Delta_+^2\xi_2^4}}
\Biggr) . \notag
\end{align}
This is again effectively a quadratic equation in $\xi_2^2$, which can be solved to yield
\begin{align}\label{eq:corPoint}
\xi_2^2 &= \frac{1}{2(1 - \abs{\Delta_0})}\biggl[(1 - \abs{\Delta_0})^2 + \Delta_-^2 - \Delta_+^2 \pm {} \\
&\eqspace
\sqrt{((1 - \abs{\Delta_0})^2 + \Delta_-^2 - \Delta_+^2)_{\vphantom{1}}^2 - 4(1 - \abs{\Delta_0})_{\vphantom{1}}^2\Delta_-^2}\biggr] . \notag
\end{align}
We should again choose the largest root, which in this case is the one with the $+$ sign, because $1 - \abs{\Delta_0} \geq 0$.

The question is now which solution yields the global minimum? In the case of the H$_2 + 2$H, we know that the ground state 1RDM has $n_{2a} = n_{2b} = 1$, so $\Delta_+ = 0$ and $-1/2 \leq \Delta_0 = \Delta_- \leq 1/2$. This greatly simplifies the problem, since $\xi$ is now directly related by the constraint~\eqref{eq:redConstr} to $\xi_2$ as
\begin{align}
\xi_1 &= \frac{-1}{\xi_2}\sqrt{2(1 - \abs{\Delta_0})\xi_2^2 - \xi_2^4 - \Delta_0^2} \\
\intertext{and the last line of the interaction energy~\eqref{eq:redInter} reduces to}
\tilde{W} &= \xi_1\xi_2 w_+ {} \\
&= -w_+\sqrt{2(1 - \abs{\Delta_0})\xi_2^2 - \xi_2^4 - \Delta_0^2} . \notag
\end{align}
Minimisation over $\xi_2$ is now straightforward and yields
\begin{equation}
\xi_2 = -\sqrt{1 - \abs{\Delta_0}} , \\
\end{equation}
The only stationary point which converges to this point is~\eqref{eq:corPoint} with the plus sign, so we can hope that this point always yields the minimum. This suspicion has been confirmed by a numerical check with \textsc{Mathematica} where the parameters are constrained as
\begin{subequations}
\begin{gather}
0 \leq \tfrac{w_-}{w_+} \leq 1 , \\
-1 \leq \Delta_\pm \leq 1 , \\
0 \leq \abs{\Delta_0} \leq 1 , \\
\label{eq:DeltaSumConstr}
\abs{\Delta_0} + \abs{\Delta_-} + \abs{\Delta_+} \leq 1 ,
\end{gather}
\end{subequations}
where the latter condition is a result of working out
\begin{subequations}
\begin{align}
\Delta_0 + \Delta_+ + \Delta_- &= n_{1a} - 1 , \\
\Delta_0 + \Delta_+ - \Delta_- &= 1 - n_{2b} , \\
\Delta_0 - \Delta_+ + \Delta_- &= 1 - n_{1b} , \\
\Delta_0 - \Delta_+ - \Delta_- &= n_{2a} - 1 .
\end{align}
\end{subequations}
By putting the most extreme occupation numbers (0 or 2) we find that all left-hand sides $\in [-1,1]$, so~\eqref{eq:DeltaSumConstr} follows.


\section{The $C_{2}$ symmetry-restricted functional for the H$_4$ systems in minimal basis}
\label{ap:paramC2v}
Though the trapezoid configuration has $C_{2v}$ as its highest point group symmetry, only one non-trivial symmetry element is relevant, since only the $1s$-orbitals are considered in the planar configuration. Here we choose to retain the rotation around an axis by $180^{\circ}$ as the non-trivial symmetry operation, so we construct a $C_{2}$ symmetry-restricted functional. Note that the groups $C_s$ and $C_i$ are isomorphic to $C_2$, so we effectively also obtain the symmetry-restricted functional for those groups (only the irrelevant relabelling $\{a,b\} \to \{a',a''\}$ or $\{a_g, a_u\}$ might be considered). To limit the discussion, we only demonstrate the construction for the $^1A$ irrep, since the construction for the $^1B$ will be analogous.

For brevity, we label symmetry adapted and orthonormalised spatial orbitals as
\begin{equation}
\{1_a, 2_a, 1_b, 2_b\} .
\end{equation}
We can use these symmetry adapted orbitals to construct the following full CI expansion for a general $^1A$ state
\begin{align}\label{eq:PsiA}
\Psi^{^1A} ={}&
c^{ab}_{11}\abs[\big]{1_a^21_b^2} + c^{aa}_{12}\abs[\big]{1_a^22_a^2} +
c^{ab}_{12}\abs[\big]{1_a^22_b^2} + {} \notag \\
&c^{ab}_{21}\abs[\big]{2_a^21_b^2} + c^{ab}_{22}\abs[\big]{2_a^22_b^2} + 
c^{bb}_{12}\abs[\big]{1_b^22_b^2} + {} \notag \\
&c^a_1\abs[\big]{1_a^21_b2_b[\alpha\beta - \beta\alpha]} +
c^a_2\abs[\big]{2_a^21_b2_b[\alpha\beta - \beta\alpha]} + {} \notag \\
&c^b_1\abs[\big]{1_b^21_a2_a[\alpha\beta - \beta\alpha]} +
c^b_2\abs[\big]{2_b^21_a2_a[\alpha\beta - \beta\alpha]} + {} \notag \\
&c^m_1\abs[\big]{1_a2_a1_b2_b[\alpha\beta\alpha\beta + \alpha\beta\beta\alpha + {} \notag \\*
&\qquad
\beta\alpha\alpha\beta + \beta\alpha\beta\alpha - 2(\alpha\alpha\beta\beta + \beta\beta\alpha\alpha)]} + {}\notag \\
&c^m_2\abs[\big]{1_a2_a1_b2_b[\alpha\beta\alpha\beta - \alpha\beta\beta\alpha - {} \notag \\*
&\qquad
\beta\alpha\alpha\beta + \beta\alpha\beta\alpha]} .
\end{align}
Since there are 12 coefficients and 6 constraints, we expect to need 6 parameters except for additional possible phase factors.

The conditions from the (spin-integrated) occupation numbers are
\begin{subequations}
\begin{align}
n^a_1 ={}& 2\bigl(\abs{c^{ab}_{11}}^2 + \abs{c^{aa}_{12}}^2 + \abs{c^{ab}_{12}}^2 + 2\abs{c^a_1}^2 + {} \notag \\*
&\hphantom{2\bigl(} \abs{c^b_1}^2 + \abs{c^b_2}^2 + 6\abs{c^m_1}^2 + 2\abs{c^m_2}^2\bigr) , \\
n^a_2 ={}& 2\bigl(\abs{c^{aa}_{12}}^2 + \abs{c^{ab}_{21}}^2 + \abs{c^{ab}_{22}}^2 + 2\abs{c^a_2}^2 + {} \notag \\*
&\hphantom{2\bigl(} \abs{c^b_1}^2 + \abs{c^b_2}^2 + 6\abs{c^m_1}^2 + 2\abs{c^m_2}^2\bigr) , \\
n^b_1 ={}& 2\bigl(\abs{c^{ab}_{11}}^2 + \abs{c^{ab}_{21}}^2 + \abs{c^{bb}_{12}}^2 + 2\abs{c^b_1}^2 + {} \notag \\*
&\hphantom{2\bigl(} \abs{c^a_1}^2 + \abs{c^a_2}^2 + 6\abs{c^m_1}^2 + 2\abs{c^m_2}^2\bigr) , \\
n^b_2 ={}& 2\bigl(\abs{c^{ab}_{12}}^2 + \abs{c^{bb}_{12}}^2 + \abs{c^{ab}_{22}}^2 + 2\abs{c^b_2}^2 + {} \notag \\*
&\hphantom{2\bigl(} \abs{c^a_1}^2 + \abs{c^a_2}^2 + 6\abs{c^m_1}^2 + 2\abs{c^m_2}^2\bigr) .
\end{align}
\end{subequations}
Taking the difference between the occupation numbers in each irrep yields
\begin{subequations}
\begin{align}
\label{eq:na1-na2}
n^a_1 - n^a_2 ={}&
2\bigl(\abs{c^{ab}_{11}}^2 + \abs{c^{ab}_{12}}^2 - \abs{c^{ab}_{21}}^2 - \abs{c^{ab}_{22}}^2 + {} \notag \\*
&\hphantom{2\bigl(}2(\abs{c^a_1}^2 - \abs{c^a_2}^2)\bigr) , \\
\label{eq:nb1-nb2}
n^b_1 - n^b_2 ={}&
2\bigl(\abs{c^{ab}_{11}}^2 - \abs{c^{ab}_{12}}^2 + \abs{c^{ab}_{21}}^2 - \abs{c^{ab}_{22}}^2 + {} \notag \\*
&\hphantom{2\bigl(}2(\abs{c^b_1}^2 - \abs{c^b_2}^2) \bigr) .
\end{align}
Assuming $c^{ab}_{11}$, $c^{ab}_{12}$, $c^{ab}_{21}$ and $c^{ab}_{22}$ to be given, $\abs{c^a_1}^2 - \abs{c^a_2}^2$ and $\abs{c^b_1}^2 - \abs{c^b_2}^2$ can be calculated.

There is one more non-trivial condition (i.e.\ apart from the trivial normalisation constraint)
\begin{equation}
n^a_1 + n^a_2 - n^b_1 - n^b_2 =
4\bigl(\abs{c^{aa}_{12}}^2 - \abs{c^{bb}_{12}}^2\bigr) .
\end{equation}
\end{subequations}
Given $c^{aa}_{12} + c^{bb}_{12}$, we can obtain $c^{aa}_{12} - c^{bb}_{12}$ or vice versa.

Now let us consider the two diagonality conditions (from the $a_1$ and $b_1$ irrep respectively)
\begin{subequations}
\begin{align}
0 ={}
&c^b_1(c^{ab}_{11} + c^{ab}_{21}) + c^b_2(c^{ab}_{12} + c^{ab}_{22}) + 2(c^a_1 + c^a_2)c^m_2 , \\
0 ={}
&c^a_1(c^{ab}_{11} + c^{ab}_{12}) + c^a_2(c^{ab}_{21} + c^{ab}_{22}) + 2(c^b_1 + c^b_2)c^m_2 .
\end{align}
\end{subequations}
Multiplying these equations by $c^b_1 + c^b_2$ and $c^a_1 + c^a_2$ respectively and subtracting them to eliminate $c^m_2$ yields
\begin{align}
&c^b_1(c^b_1 + c^b_2)(c^{ab}_{11} + c^{ab}_{21}) + c^b_2(c^b_1 + c^b_2)(c^{ab}_{12} + c^{ab}_{22}) \\
&\quad = c^a_1(c^a_1 + c^a_2)(c^{ab}_{11} + c^{ab}_{12}) + c^a_2(c^a_1 + c^a_2)(c^{ab}_{21} + c^{ab}_{22}) . \notag 
\end{align}
Now writing everything in terms of $c^b_1 \pm c^b_2$ and $c^a_1 \pm c^a_2$ we obtain
\begin{align}
(c^b_1 + c^b_2)&\bigl[(c^b_1 + c^b_2)(c^{ab}_{11} + c^{ab}_{12} + c^{ab}_{21} + c^{ab}_{22}) + {} \notag \\*
&\hphantom{\bigl[}
(c^b_1 - c^b_2)(c^{ab}_{11} - c^{ab}_{12} + c^{ab}_{21} - c^{ab}_{22})\bigr] \\
={}&
(c^a_1 + c^a_2)\bigl[(c^a_1 + c^a_2)(c^{ab}_{11} + c^{ab}_{12} + c^{ab}_{21} + c^{ab}_{22}) + {} \notag \\*
&\hphantom{(c^a_1 + c^a_2)\bigl[}
(c^a_1 - c^a_2)(c^{ab}_{11} + c^{ab}_{12} - c^{ab}_{21} - c^{ab}_{22})\bigr] \notag ,
\end{align}
which can be rearranged as
\begin{multline}
(c^{ab}_{11} + c^{ab}_{12} + c^{ab}_{21} + c^{ab}_{22})\bigl[(c^a_1 + c^a_2)^2 - (c^b_1 + c^b_2)^2\bigr] \\
\begin{aligned}
{} ={}& (\abs{c^b_1}^2 - \abs{c^b_2}^2)(c^{ab}_{11} - c^{ab}_{12} + c^{ab}_{21} - c^{ab}_{22}) - {} \\
&(\abs{c^a_1}^2 - \abs{c^a_2}^2)(c^{ab}_{11} + c^{ab}_{12} - c^{ab}_{21} - c^{ab}_{22}) .
\end{aligned}
\end{multline}
Since $\abs{c^a_1}^2 - \abs{c^a_2}^2$ and $\abs{c^b_1}^2 - \abs{c^b_2}^2$ are known from~\eqref{eq:na1-na2} and~\eqref{eq:nb1-nb2} respectively, we have an equation for $(c^a_1 + c^a_2)^2 - (c^b_1 + c^b_2)^2$.

We have now all the ingredients for a parametrisation. As parameters we choose the following quantities
\begin{subequations}
\begin{align}
\xi_1 &= c^{ab}_{11} , \\
\xi_2 &= c^{ab}_{12} , \\
\xi_3 &= c^{ab}_{21} , \\
\xi_4 &= c^{ab}_{22} , \\
\xi_5 &= c^{aa}_{12} + c^{bb}_{12} , \\
\xi_6 &= c^a_1 + c^a_2 + c^b_1 + c^b_2 .
\end{align}
\end{subequations}
In a first step we can calculate the following intermediates
\begin{subequations}
\begin{align}
\zeta_3 &= \abs{c^a_1}^2 - \abs{c^a_2}^2 \\
&= \thalf\bigl(n^a_1 - n^a_2\bigr) -
\abs{c^{ab}_{11}}^2 - \abs{c^{ab}_{12}}^2 + \abs{c^{ab}_{21}}^2 + \abs{c^{ab}_{22}}^2 , \notag \\
\zeta_4 &=\abs{c^b_1}^2 - \abs{c^b_2}^2 \\
&= \thalf\bigl(n^b_1 - n^b_2\bigr) -
\abs{c^{ab}_{11}}^2 + \abs{c^{ab}_{12}}^2 - \abs{c^{ab}_{21}}^2 + \abs{c^{ab}_{22}}^2 , \notag \\
\zeta_5 &= c^{aa}_{12} - c^{bb}_{12}
= \frac{\abs{c^{aa}_{12}}^2 - \abs{c^{bb}_{12}}^2}{4\xi_5} .
\end{align}
Then we evaluate
\begin{align}
&\zeta_6 = c^a_1 + c^a_2 - c^b_1 - c^b_2 \\
&= \frac{\zeta_4(c^{ab}_{11} - c^{ab}_{12} + c^{ab}_{21} - c^{ab}_{22}) -
\zeta_3(c^{ab}_{11} + c^{ab}_{12} - c^{ab}_{21} - c^{ab}_{22})}
{\xi_6(c^{ab}_{11} + c^{ab}_{12} + c^{ab}_{21} + c^{ab}_{22})} \notag 
\end{align}
and subsequently
\begin{align}
\zeta_7 &= c^a_1 + c^a_2 = (\xi_6 + \zeta_6)/2 , \\
\zeta_8 &= c^b_1 + c^b_2 = (\xi_6 - \zeta_6)/2 , \\
\zeta_9 &= c^a_1 - c^a_2 = \zeta_3/\zeta_7 , \\
\zeta_{10} &= c^b_1 - c^b_2 = \zeta_4/\zeta_8 .
\end{align}
\end{subequations}
The wavefunction coefficients are now obtained as
\begin{subequations}
\begin{align}
c^{ab}_{11} &= \xi_1 , \\
c^{ab}_{12} &= \xi_2 , \\
c^{ab}_{21} &= \xi_3 , \\
c^{ab}_{22} &= \xi_4 , \\
c^{aa}_{12} &= (\xi_5 + \zeta_5)/2 , \\
c^{bb}_{12} &= (\xi_5 - \zeta_5)/2 , \\
c^a_1 &= (\zeta_7 + \zeta_9)/2 , \\
c^a_2 &= (\zeta_7 - \zeta_9)/2 , \\
c^b_1 &= (\zeta_8 + \zeta_{10})/2 , \\
c^b_2 &= (\zeta_8 - \zeta_{10})/2 , \\
c^m_2 &= \frac{c^b_1(c^{ab}_{11} + c^{ab}_{21}) + c^b_2(c^{ab}_{12} + c^{ab}_{22})}{2\zeta_7} , \\
c^m_1 &= \tfrac{1}{12}
\bigl(1 - \xi_1^2 - \xi_2^2 - \xi_3^2 - \xi_4^2 - \abs{c^{aa}_{12}}^2 - \abs{c^{bb}_{12}}^2 - {} \notag \\*
&\eqspace\hphantom{\tfrac{1}{12}\bigl(}
\zeta_7^2 - \zeta_8^2 - \zeta_9^2 - \zeta_{10}^2 - 4\abs{c^m_2}^2\bigr)^{1/2} .
\end{align}
\end{subequations}

\bibliography{bible}

\end{document}